\documentclass[12pt, leqno]{article}
\usepackage{amsmath,caption,setspace,multirow}
\usepackage[top=1.12in, bottom=1.12in, left=1.12in, right=1.12in]{geometry}

\usepackage[round]{natbib}
\usepackage{color,soul}

\DeclareCaptionStyle{italic}[justification=centering]{labelfont={bf},textfont={it},labelsep=colon}
\usepackage{graphicx,psfrag,epsf,textcomp,epstopdf}
\usepackage{enumerate, dsfont}
\usepackage{natbib}
\usepackage{url,xcolor}
\usepackage{booktabs, subfig, bm, paralist,mathpazo,tikz,todonotes,longtable,microtype}
\usepackage[]{algorithm2e}

\usepackage[pdftex,colorlinks=true]{hyperref}
\definecolor{darkblue}{rgb}{0,0,.6}
\hypersetup{citecolor=darkblue,linkcolor=darkblue,urlcolor=darkblue}
\definecolor{DarkRed}{rgb}{.7,0,.4}

\newcommand\Tau{\mathcal{T}}
\usepackage{comment}

\newcommand{\blind}{0}

\newcommand{\X}{\mathcal{X}}
\newcommand{\Y}{\mathcal{Y}}

\graphicspath{{plots_separate_figures/}}

\newsavebox\CBox

 \newtheorem{@definition}{\sc Definition}[section]

  \renewcommand\X{\mathcal{X}}

\begin{document}

\def\spacingset#1{\renewcommand{\baselinestretch}{#1}\small\normalsize} \spacingset{1}

\if0\blind
{
  \title{\bf On function-on-function regression: \mbox{Partial least squares approach}}
    \author{
Ufuk Beyaztas \\
Department of Statistics \\
Bartin University \\
\\
Han Lin Shang\footnote{Corresponding author: Research School of Finance, Actuarial Studies, and Statistics, Australian National University, Level 4, Building 26C, Kingsley Street, Acton, ACT 2601, Australia; Telephone: +61(2) 61250535; Fax: +61(2) 61250087} \\
{Research School of Finance, Actuarial Studies, and Statistics} \\
{Australian National University}
 }
  \maketitle
} \fi

\if1\blind
{
  \bigskip
  \bigskip
  \bigskip
  \begin{center}
    {\LARGE\bf On function-on-function regression: \mbox{Partial least squares approach}}
\end{center}
  \medskip
} \fi

\maketitle

\begin{abstract}
Functional data analysis tools, such as function-on-function regression models, have received considerable attention in various scientific fields because of their observed high-dimensional and complex data structures. Several statistical procedures, including least squares, maximum likelihood, and maximum penalized likelihood, have been proposed to estimate such function-on-function regression models. However, these estimation techniques produce unstable estimates in the case of degenerate functional data or are computationally intensive. To overcome these issues, we proposed a partial least squares approach to estimate the model parameters in the function-on-function regression model. In the proposed method, the $B$-spline basis functions are utilized to convert discretely observed data into their functional forms. Generalized cross-validation is used to control the degrees of roughness. The finite-sample performance of the proposed method was evaluated using several Monte-Carlo simulations and an empirical data analysis. The results reveal that the proposed method competes favorably with existing estimation techniques and some other available function-on-function regression models, with significantly shorter computational time.
\end{abstract}


\newpage
\spacingset{1.56}

\section{Introduction\label{sec:intro}}

Recent advances in computer storage and data collection have enabled researchers in diverse branches of science such as, for instance, chemometrics, meteorology, medicine, and finance, recording data of characteristics varying over a continuum (time, space, depth, wavelength, etc.). Given the complex nature of such data collection tools, the availability of functional data, in which observations are sampled over a fine grid, has progressively increased. Consequently, the interest in functional data analysis (FDA) tools is significantly increasing over the years. \cite{ramsay2002, ramsay2006}, \cite{ferraty2006}, \cite{horvath2012} and \cite{cuevas2014} provide excellent overviews of the research on theoretical developments and case studies of FDA tools.

Functional regression models in which both the response and predictors consist of curves known as, function-on-function regression, have received considerable attention in the literature. The main goal of these regression models is to explore the associations between the functional response and the functional predictors observed on the same or potentially different domains as the response function. In this context, two key models have been considered: the \textit{varying-coefficient} model and the \textit{function-on-function regression model} (FFRM). The varying-coefficient model assumes that the functional response $\Y(t)$ and functional predictors $\X(t)$ are observed in the same domain. Its estimation and test procedures  have been studied by numerous authors, including \cite{fan1999}, \cite{Hoover1998}, \cite{Brumback1998}, \cite{WuChiang2000}, \cite{HWZ2002, HWZ2004}, \cite{senturk2005}, \cite{cardotSarda2008}, \cite{WFM2010} and \cite{ZFK2014}  among many others. In contrast, the FFRM considers cases in which the functional response $\Y(t)$ for a given continuum $t$ depends on the full trajectory of the predictors $\X(s)$. Compared with the varying-coefficient model, the FFRM is more natural; therefore, we restrict our attention to the FFRM for this study.

The FFRM was first proposed by \cite{ramsay1991}, who extended the traditional multivariate regression model to the infinite-dimensional case. In the FFRM, the association between the functional response and the functional predictors is expressed by integrating the full functional predictor weighted by an unknown bivariate coefficient function. More precisely, if $\Y_i(t)$ ($i = 1, \cdots, N$) and $\X_{im}(s)$ ($m = 1, \cdots, M$), respectively, denote a set of functional responses and $M$ sets of functional predictors with $s \in \left[0, S\right]$ and $t \in \left[0, \Tau\right]$, where $S$ and $\Tau$ are closed and bounded intervals on the real line, then the FFRM for $\Y_i(t)$ and $\X_{im}(s)$ is constructed as follows:
\begin{equation}
\Y_i(t) = \beta_0(t) + \sum_{m=1}^M \int_{S} \X_{im}(s) \beta_m(s,t) ds + \epsilon_i(t), \label{Eq:FLM}
\end{equation}
where $\beta_0(t)$ is the mean response function, $\beta_m(s,t)$ is the bivariate coefficient function, and $\epsilon_i(t)$ denotes an independent random error function having a normal distribution with mean vector $\mathbf{0}$ and variance-covariance matrix $\mathbf{\Sigma}_{\epsilon}$, i.e., $\epsilon_i(t) \sim \text{N}(\mathbf{0}, \mathbf{\Sigma}_{\epsilon})$. The main purpose of model~\eqref{Eq:FLM} is to estimate the bivariate coefficient function $\beta_m(s,t)$. In this context, \cite{yamanishi2003} proposed a geographically weighted regression model to explore the functional relationship between the variables; \cite{ramsay2006} proposed a least squares (LS) method to estimate $\beta_m(s,t)$ by minimizing the integrated sum of squares; \cite{yao2005} extended the FFRM to the analysis of sparse longitudinal data and discussed the estimation procedures;  \cite{MullerYao2008} proposed a functional additive regression model where regression parameters are estimated using regularization; \cite{matsui2009} suggested a maximum penalized likelihood (MPL) approach to estimate the coefficient function in the FFRM; \cite{wang2014} proposed a linear mixed regression model and estimated the model parameters via the expectation/conditional maximization either algorithm; \cite{ivanescu2015} developed several penalized spline approaches to estimate the FFRM parameters using the mixed model representation of penalized regression; and \cite{chiou2016} proposed a multivariate functional regression model to analyze multivariate functional data.

Most investigations on parameter estimation in FFRM have generally focused on the LS, maximum likelihood (ML), and MPL approaches. While these approaches work well in certain circumstances, they are characterized by several drawbacks. For instance, the ML and LS methods produce unstable estimates when functional data have degenerate structures \citep[see][]{matsui2009}. They also encounter a singular matrix problem when a large number of functional predictors are included in the FFRM. Alternatively, the singular matrix problem can also occur when a large number of basis functions are used to approximate those functions. In such cases, the LS and ML methods typically fail to provide an estimate for $\beta_m(s,t)$. Although the MPL method can overcome such difficulties and produce consistent estimates, it is computationally time-consuming for a computer with standard memory. It may not be possible to obtain the MPL estimates where a large number of basis functions are used to approximate the functional data. In this paper, we propose a partial least squares (PLS) approach to estimate the parameter of the FFRM to overcome these vexing issues. 

The functional counterparts of the PLS method, when the functional data consist of scalar response and functional predictors, were proposed by \cite{PredSap}, \cite{ResiOgden}, \cite{Kramer}, and \cite{Agu2010}. \cite{Febrero} compared these methods and discussed their advantages and disadvantages. \cite{HydSh2009} proposed a weighted functional partial least squares regression method for forecasting functional time series. Their method is based on a lagged functional predictor and a functional response. In this paper, we proposed an extended version of the functional partial least squares regression (FPLSR) of \cite{PredSc}. The proposed method differs from the previous FPLSR in two respects. First, while the FPLSR considers only one functional predictor in the model, our approach allows for more than one functional predictor. Second, the FPLSR uses a fixed smoothing parameter when converting the discretely observed data to functional form. However, our approach uses a grid search to determine the optimal smoothing parameter.

In summary, our proposed method works as follows. First, the $B$-spline basis function expansion is used to express discretely observed data as smooth functions. The number of basis functions is determined using the penalized LS, and the smoothing parameter that controls the roughness of the expansion is specified by the generalized cross-validation (GCV). The discretized version of the smooth coefficient function obtained by the basis function expansion is solved for a matrix (say $\mathbf{B}$ in~\eqref{eq:Regr}) using a PLS algorithm. In this study, we used the two fundamental PLS algorithms found in the literature to estimate $\mathbf{B}$ - nonlinear iterative partial least squares (NIPALS) \citep{nipals} and simple partial least squares (SIMPLS) \citep{simpls}. Finally, the estimate of the coefficient function $\beta_m(s,t)$ was obtained by applying the smoothing step. The main advantage of the proposed method is that it bypasses the singular matrix problem. Further, the proposed method increases the predicting accuracy of the FFRM and is more efficient compared with some other available estimation methods.

The remainder of this paper proceeds as follows. Section~\ref{sec:meth} is dedicated to the methodology of the proposed method. Section \ref{sec:sim} evaluates the finite-sample performance of the proposed method using several Monte-Carlo experiments. Section~\ref{sec:real} applies the proposed method to a dataset on solar radiation prediction. Section~\ref{sec:conc} concludes the paper and provides several future research directions.

\section{Methodology}\label{sec:meth} 

For the FFRM provided by~\eqref{Eq:FLM}, the functional random variables are assumed to be an element of $\mathcal{L}_2$, which expresses square-integrable and real-valued functions. They are further assumed to be second-order stochastic processes with finite second-order moments. The association between these functional variables is characterized by the surface $\beta_m(s,t) \in \mathcal{L}_2$, where $\mathcal{L}_2$ denotes a square-integrable functional space. Without loss of generality, the mean response function $\beta_0(t)$ is eliminated from the model~\eqref{Eq:FLM} by centering the functional response and functional predictor variables.

If $\Y^*_i(t) = \Y_i(t) - \overline{\Y}(t)$, $\X^*_{im}(s) = \X_{im}(s) - \overline{\X}_m(s)$ and $\epsilon^*_i(t) = \epsilon_i(t) - \overline{\epsilon}(t)$ are used to denote the centered versions of the functional variables and the error function defined in~\eqref{Eq:FLM}, the model~\eqref{Eq:FLM} can be re-expressed as follows:
\begin{equation}
\Y^*_i(t) = \sum_{m=1}^M \int_{S} \X^*_{im}(s) \beta_m(s,t) ds + \epsilon^*_i(t). \label{eq:regc}
\end{equation}
By custom, we expressed the functional variables and the bivariate coefficient function as basis function expansions before fitting the FFRM.

Initially, let $x(t)$ denote a function finely sampled on a grid $t \in [0, \Tau]$. Based on a pre-determined basis and a sufficiently large number of basis functions $K$, it can be approximated as $ x(t) \approx \sum_{k=1}^K c_k \phi_k(t)$, where $\phi_k(t)$ and $c_k$, for $k = 1, \cdots, K$, represent the $k^{th}$ basis function and its associated coefficient vector, respectively. In this study, the functions were approximated using $B$-spline basis and the number of basis functions were determined according to GCV. Similarly, the (centered) functional variables and the bivariate coefficient function in~\eqref{eq:regc} can be written as basis function expansions as follows:
\begin{align}
\Y^*_i(t) &= \sum_{k=1}^{K_{\Y}} c_{ik} \phi_k(t) = \mathbf{c}_i \mathbf{\Phi}(t) \qquad \forall t \in \Tau, \label{eq:Ybas} \\
\X^*_{im}(s) &= \sum_{j=1}^{K_{m,\X}} d_{imj} \psi_{mj}(s) = \mathbf{d}_{im} \mathbf{\Psi}(s) \qquad \forall s \in S, \label{eq:Xbas} \\
\beta_m(s,t) &= \sum_{j,k} \psi_{mj}(s) b_{mjk} \phi_k(t) = \mathbf{\Psi}_m(s) \mathbf{B}_m \mathbf{\Phi}(t) \qquad \forall t\in \Tau, ~ \forall s \in S, \label{eq:Bbas}
\end{align}
where $\mathbf{\Phi}(t)$ and $\mathbf{\Psi}(s)$ are the vectors of the basis functions with dimensions $K_{\Y}$ and $K_{m,\X}$, respectively, $\mathbf{c}_{i}$ and $\mathbf{d}_{im}$, respectively, are the $K_{\Y}$ and $K_{m,\X}$ dimensional coefficient vectors, and $\mathbf{B}_m$ is a $K_{m,\X} \times K_{\Y}$ dimensional coefficient matrix. Replacing~\eqref{eq:Ybas} to~\eqref{eq:Bbas} with~\eqref{eq:regc} yields:
\begin{align}\label{regs}
\mathbf{c}_i \mathbf{\Phi}(t) &= \sum_{m=1}^M \mathbf{d}_{im} \bm{\zeta}_{\psi_m} \mathbf{B}_m \mathbf{\Phi}(t) + \epsilon^*_i(t), \nonumber \\
&= z_i \mathbf{B} \mathbf{\Phi}(t) + \epsilon^*_i(t),
\end{align}
where $\bm{\zeta}_{\psi_m} = \int_{S} \psi_m(s) \psi^\top_m(s) ds$ is a $K_{m,\X} \times K_{m,\X}$ cross-product matrix, $z_i = \left( \mathbf{d}^\top_{i1} \bm{\zeta}_{\psi_1}, \cdots, \mathbf{d}^\top_{iM} \bm{\zeta}_{\psi_M} \right)^\top$ is a vector of dimension $\sum_{m=1}^M K_{m,\X}$, and $\mathbf{B} = \left( \mathbf{B}_1, \cdots, \mathbf{B}_M \right)^\top$ is the coefficient matrix with dimensions $\sum_{m=1}^M K_{m,\X} \times K_{\Y}$. Let $\mathbf{C} = \left( \mathbf{c}_1, \cdots, \mathbf{c}_N \right)^\top$, $\mathbf{Z} = \left( \mathbf{z}_1, \cdots, \mathbf{z}_N \right)^\top$ and $\pmb{\varepsilon}(t) = \left( \epsilon^*_1(t), \cdots, \epsilon^*_N(t) \right)^\top$, the model~\eqref{regs} can then be rewritten as follows:
\begin{equation}\label{eq:Regr}
\mathbf{C} \mathbf{\Phi}(t) = \mathbf{Z} \mathbf{B} \mathbf{\Phi}(t) + \pmb{\varepsilon}(t).
\end{equation}

Assuming that the error function $\pmb{\varepsilon}(t)$ in~\eqref{eq:Regr} can also be represented as a basis function expansion, then  $\pmb{\varepsilon}(t) = \mathbf{e} \mathbf{\Phi}(t)$ with $\mathbf{e} = \left( \mathbf{e}_{1}, \cdots, \mathbf{e}_{N} \right)^\top$, where each $\mathbf{e}_i$ consists of independently and identically distributed (iid) random variables $\mathbf{e}_{i} = \left( e_{i1}, \cdots, e_{iK} \right)^\top$ having a normal distribution with mean $\mathbf{0}$ and variance-covariance matrix $\mathbf{\Sigma}$. Replacing $\pmb{\varepsilon}(t)$ with $\mathbf{e} \mathbf{\Phi}(t)$ in~\eqref{eq:Regr}, and multiplying the whole equation by $\mathbf{\Phi}^\top(t)$ from the right and integrating with respect to $\Tau$, yields:
\begin{equation*}\label{eq:Reg-final}
\mathbf{C} = \mathbf{Z} \mathbf{B} + \mathbf{e}.
\end{equation*}
Estimating $\mathbf{B}$ is an ill-posed problem. The dimension of $\mathbf{B}$ increases exponentially when a large number of basis functions are used to approximate the functions or when a large number of predictors are used in the model. In such cases, traditional estimation methods such as LS and ML fail to provide an estimate for $\mathbf{B}$. However, the MPL method can produce a stable estimate for $\mathbf{B}$ as long as the functional data are approximated by a small number of basis functions. Because it is computationally intensive, obtaining an MPL estimate of $\mathbf{B}$ may not be possible. This is the case when a relatively large number of basis functions are used to convert discretely observed data into the functional form. In this paper, we propose using the PLS approach to obtain a stable estimate for $\mathbf{B}$. Compared with MPL, PLS has several important advantages, including flexibility, straightforward interpretation, and fast computation ability in high-dimensional settings. Note that our proposal is based on an extended version of the FPLSR suggested by \cite{PredSc}.

\subsection{PLS for the function-on-function regression model}
Let $\pmb{\X}^*(s) = \left( \X^*_1(s), \cdots, \X^*_M(s) \right)$ with $\X^*_m(s) = \left(\X^*_{m1}, \cdots, \X^*_{mN} \right)$ ($m = 1, \cdots, M$) and $\Y^*(t) = \left( \Y^*_1(t), \cdots, Y^*_N(t) \right)$ denote a matrix of $M$ sets of centered functional predictors of size $\left(M \times N \right) \times J_x$ and a matrix of a set of centered functional response of size $N \times J_y$, respectively. Herein, the terms $J_x$ and $J_y$ denote the lengths of time spans where the predictors and response functions observed. Let us now denote the FFRM of $\Y^*(t)$ on $\pmb{\X}^*(s)$ as follows:
\begin{equation}\label{ffrm_all}
\Y^*(t) = \int_S \pmb{\X}^*(s) \pmb{\beta}(s,t) ds + \pmb{\epsilon}^*(t),
\end{equation}
where $\pmb{\beta}(s,t)$ and $\pmb{\epsilon}^*(t)$ denote the $M$ sets of bivariate coefficient functions and error functions, respectively.
The PLS components of the FFRM~\eqref{ffrm_all} may be obtained as solutions of Tucker's criterion extended to functional variables as follows:
\begin{equation*}
\underset{\begin{subarray}{c}
  \kappa \in \mathcal{L}_2,~ \Vert \kappa \Vert_{\mathcal{L}_2} = 1 \\
  \zeta \in \mathcal{L}_2,~ \Vert \zeta \Vert_{\mathcal{L}_2} = 1
  \end{subarray}}{\max} \text{Cov}^2 \left( \int_S \pmb{\X}^*(s) \kappa(s) ds, ~ \int_{\Tau} \Y^*(t) \zeta(t) dt \right).
\end{equation*}

The functional PLS components also correspond to the eigenvectors of Escoufier's operators \citep{PredSap}. Let $Z \in \mathcal{L}_2$ denote a random variable. Then, the Escoufier's operators of the centered functional response, $W^{\Y^*}$, and the matrix of $M$ sets of centered functional predictors, $W^{\pmb{\X}^*}$, are given as follows:
\begin{align*}
W^{\Y^*} &= \int_{\Tau} \mathbb{E} \left[ \Y^*(t) Z \right] \Y^*(t) dt \\
W^{\pmb{\X}^*} &= \int_S \mathbb{E} \left[ \pmb{\X}^*(s) Z \right] \pmb{\X}^*(s) ds.
\end{align*}
The first PLS component of the FFRM~\eqref{ffrm_all}, $\eta_1$, is then equal to the eigenvector of the largest eigenvalue of the product of Escoufier's operators, $\lambda$:
\begin{equation*}
W^{\pmb{\X}^*} W^{\Y^*} \eta_1 = \lambda \eta_1.
\end{equation*}
The first PLS component is defined as follows:
\begin{equation*}
\eta_1 = \int_S \kappa_1(s) \pmb{\X}^*(s) ds,
\end{equation*}
where the weight function $\kappa_1(s)$ is as follows:
\begin{equation*}
\kappa_1(s) = \frac{\int_{\Tau} \mathbb{E} \left[ \Y^*(t) \pmb{\X}^*(s)\right] dt}{\sqrt{\int_S \left( \int_{\Tau} \mathbb{E} \left[ \Y^*(t) \pmb{\X}^*(s)\right] dt \right)^2 ds}}.
\end{equation*}

The PLS approach is an iterative method, which maximizes the squared covariance between the response and predictor variables as a solution to Tucker's criterion in each iteration. Let $h = 1, 2, \cdots$ denote the iteration number. At each step $h$, the PLS components are determined by the residuals of the regression models constructed at the previous step as follows:
\begin{align*}
\pmb{\X}^*_h(s) &= \pmb{\X}^*_{h-1}(s) - p_h(s) \eta_h, \\
\Y^*_h(t) &= \Y^*_{h-1}(t) \zeta_h(t) \eta_h,
\end{align*}
where $\pmb{\X}^*_0(s) = \pmb{\X}^*(s)$, $\Y^*_0(t) = \Y^*(t)$, $p_h(s) = \frac{\mathbb{E} \left[ \pmb{\X}^*_{h-1}(s) \eta_h\right]}{\mathbb{E} \left[ \eta_h^2 \right]}$, and $\zeta_h(t) = \frac{\mathbb{E} \left[ \Y^*_h(t) \eta_h \right]}{\mathbb{E} \left[ \eta_h^2 \right]}$. Then, the $h$\textsuperscript{th} PLS component, $\eta_h$ corresponds to the eigenvector of the largest eigenvalue of the product of Escoufier's operators computed at step $h-1$ as follows:
\begin{equation*}
W_{h-1}^{\pmb{\X}^*} W_{h-1}^{\Y^*} \eta_h = \lambda \eta_h.
\end{equation*}
Similarly to the first PLS component, the  $h$\textsuperscript{th} PLS component is obtained as follows:
\begin{equation*}
\eta_h = \int_S \kappa_h(s) \pmb{\X}^*_{h-1}(s) ds,
\end{equation*}
where the weight function $\eta_h$ is given by:
\begin{equation*}
\kappa_h(s) = \frac{\int_{\Tau} \mathbb{E} \left[ \Y^*_{h-1}(t) \pmb{\X}^*_{h-1}(s)\right] dt}{\sqrt{\int_S \left( \int_{\Tau} \mathbb{E} \left[ \Y^*_{h-1}(t) \pmb{\X}^*_{h-1}(s)\right] dt \right)^2 ds}}.
\end{equation*}
Finally, the ordinary linear regressions of $\pmb{\X}^*_{h-1}(s)$ and $\Y^*_{h-1}(t)$ on $\eta_h$ are conducted to complete the PLS regression.

The observations of the functional response and functional predictors are intrinsically infinite-dimensional. However, in practice, they are observed in the sets of discrete time points. In this case, the direct estimation of a functional PLS regression becomes an ill-posed problem since the Escoufier's operators are needed to be estimated using the discretely observed observations. To overcome this problem, we consider the basis function expansions of the functional variables.

Let us now consider the basis expansions of $\Y^*(t)$ and $\pmb{\X^*}(s)$ as follows:
\begin{align*}
\Y^*(t) &= \sum_{k=1}^{K_{\Y}} C_{k} \phi_k(t) = \mathbf{C} \mathbf{\Phi}(t) \\
\pmb{\X}^*(s) &= \sum_{j=1}^{K_{\X}} \pmb{C}_j \pmb{\psi}_j(s) = \mathbf{D} \mathbf{\Psi}(s).
\end{align*}
Denote by $\pmb{\Phi} = \int_{\Tau} \pmb{\Phi}(t) \pmb{\Phi}^\top(t) dt$ and $\pmb{\Psi} = \int_S \pmb{\Psi}(s) \pmb{\Psi}^\top(s) ds$ the $K_{\Y} \times K_{\Y}$ and $K_{\pmb{\X}} \times K_{\pmb{\X}}$ dimensional symmetric matrices of the inner products of the basis functions, respectively. Also, let $\pmb{\Phi}^{1/2}$ and $\pmb{\Psi}^{1/2}$ denote the square roots of $\pmb{\Phi}$ and $\pmb{\Psi}$, respectively. Then, we consider the PLS regression of  $\pmb{C} \pmb{\Phi}^{1/2}$ on $\pmb{D} \pmb{\Psi}^{1/2}$ to approximate the PLS regression of $\Y^*(t)$ on $\pmb{\X}^*(s)$ as follows:
\begin{equation*}
\pmb{C} \pmb{\Phi}^{1/2} = \pmb{D} \pmb{\Psi}^{1/2} \pmb{\Xi} + \pmb{\delta},
\end{equation*}
where $\pmb{\Xi}$ and $\pmb{\delta}$ denote the regression coefficients and the residuals, respectively. Now let $\widehat{\pmb{\Xi}}^h$ denote the estimate of $\pmb{\Xi}$ using the PLS regression at step $h$. Then we have,
\begin{align*}
\pmb{C} \pmb{\Phi}^{1/2} &= \pmb{D} \pmb{\Psi}^{1/2} \widehat{\pmb{\Xi}}^h, \\
\widehat{\Y}^*(t) &= \int_S \pmb{\X}^*(s) \pmb{\Theta}^h(s,t) ds,
\end{align*}
where
\begin{equation*}
\pmb{\Theta}^h(s,t) = \sum_{k=1}^{K_{\Y}} \sum_{j=1}^{K_{\X}} \left( \left( \pmb{\Psi}^{1/2}\right)^{-1} \widehat{\pmb{\Xi}}^h \left( \pmb{\Phi}^{1/2}\right)^{-1} \right) \pmb{\psi}_j(s) \phi_k(t).
\end{equation*}
Herein, the term $\pmb{\Theta}^h(s,t)$ denotes the PLS approximation of the coefficient function $\pmb{\beta}(s,t)$ given in~\eqref{ffrm_all}.

Throughout this paper, two main PLS algorithms were used to obtain the model parameters: NIPALS and SIMPLS. While the NIPALS algorithm iteratively deflates the functional predictor and functional response, the SIMPLS algorithm iteratively deflates the covariance operator. In our numerical analyses, the functions \texttt{plsreg2} and \texttt{pls.regression} of the \texttt{R} packages \texttt{plsdepot} \citep{plsdepot} and \texttt{plsgenomics} \citep{plsgenomics} were used to perform NIPALS and SIMPLS algorithms, respectively.

\section{Simulation Studies}\label{sec:sim}

Various Monte-Carlo experiments were conducted under different scenarios to investigate the finite-sample performances of the proposed PLS-based methods. Throughout these experiments, $\text{MC} = 1,000$ Monte-Carlo simulations were performed, and the results were compared with LS, ML, MPL, and two available FFRM models: 
\begin{inparaenum}
\item[1)] penalized flexible functional regression (PFFR) from \cite{ivanescu2015} \citep[refer to the \texttt{R} package ``\texttt{refund}'' from][for details]{refund} and
\item[2)] the functional regression with functional response (FREG) from \cite{ramsay2006} \citep[refer to the \texttt{R} package ``\texttt{fda.usc}'' from][for details]{fdausc}. 
\end{inparaenum}

Throughout the experiments, the following simple FFRM was considered:
\begin{equation*}
\Y_i(t) = \beta_0(t) + \int_{\Tau} \X_i(s) \beta_1(s,t) ds + \epsilon_i(t),
\end{equation*}
where $s \in S$, $t \in \Tau$, and $N = 100$ and $200$ individuals were considered. A comparison was made using the average mean squared error (AMSE). For each experiment, the generated data were divided into two parts: 
\begin{inparaenum}
\item[1)] The first half of the data were used to build the FFRM, and the following AMSE was calculated: 
\begin{equation*}
\text{AMSE} = (N/2)^{-1} \sum_{i=1}^{N/2} \left[ \Y_i(t) - \widehat{\Y}_i(t) \right]^2, 
\end{equation*}
where $\widehat{\Y}_i(t)$ denotes the fitted function for $i$\textsuperscript{th} individual.
\item[2)] The second part of the data was used to evaluate the prediction performances of the methods based on the constructed FFRMs using the first-half of the data:
\begin{equation*}
\text{AMSE}_p = (N/2)^{-1} \sum_{i=N/2+1}^N \left[ \Y_i(t) - \widehat{\Y}^*_i(t) \right]^2, 
\end{equation*}
where $\widehat{\Y}^*_i(t)$ denotes the predicted response function for $i$\textsuperscript{th} individual.
\end{inparaenum}
Also, we applied the model confidence set (MCS) procedure proposed by \cite{Hansen} \citep[refer to the \texttt{R} package ``\texttt{MCS}'' from][for details]{MCS} on the prediction errors obtained by the FFRM procedures to  determine superior method(s). The MCS procedure was performed using 5,000 bootstrap replications at a 95\% confidence level. Computations were performed using \cite{Team19} on an Intel Core i7 6700HQ 2.6 GHz PC.

The following process was used to generate functional variables:
\begin{itemize}
\item Generate the observations of the predictor variable $\X$ at discrete time points $s_j$ as follows:
\begin{equation*}
\X_{ij} = \kappa_i(s_j) + \epsilon_{ij},
\end{equation*}
where $j = 1, \cdots, 50$, $\epsilon_{ij} \sim N(0, 1)$, $s_j \sim U(-1, 1)$, and $\kappa_i(s)$ is generated as:
\begin{equation*}
\kappa_i(s) = \cos \left[ \exp \left( a_{1_i} s\right) \right] + a_{2_i} s,
\end{equation*}
where $a_{1_i} \sim \text{N}(2, 0.02^2)$ and $a_{2_i} \sim \text{N}(-3, 0.04^2)$.
\item Similarly, generate the data points of the response variable $\Y$ at time points $t_j$ using the following process:
\begin{equation*}
\Y_{ij} = \eta_i(t_j) + \epsilon_{ij},
\end{equation*}
where $t_j \sim U(-1, 1)$ and $\eta_i(t)$ is generated as:
\begin{equation*}
\eta_i(t) = \vartheta_i(t) + \varepsilon_i(t),
\end{equation*}
where $\vartheta_i(t) = \sin \left[ \exp \left( a_{1_i} t \right) \right] + a_{2_i} t + 2 t^2$, $\varepsilon_i(t) = \pmb{e}^\top_i \pmb{\Phi}(t)$, $\pmb{e}_i$s are iid multivariate Gaussian random errors with mean $\mathbf{0}$ and variance-covariance matrix $\mathbf{\Sigma} = [(0.5^{\vert k - l \vert}) \rho ]_{k,l}$, and $\pmb{\Phi}(t)$ is the $B$-spline basis function. Throughout the simulations, four different variance parameters were considered: $\rho = [0.5, 1, 2, 4]$.
\end{itemize}

\begin{figure}[!htbp]
  \centering
  \includegraphics[width=5.9cm]{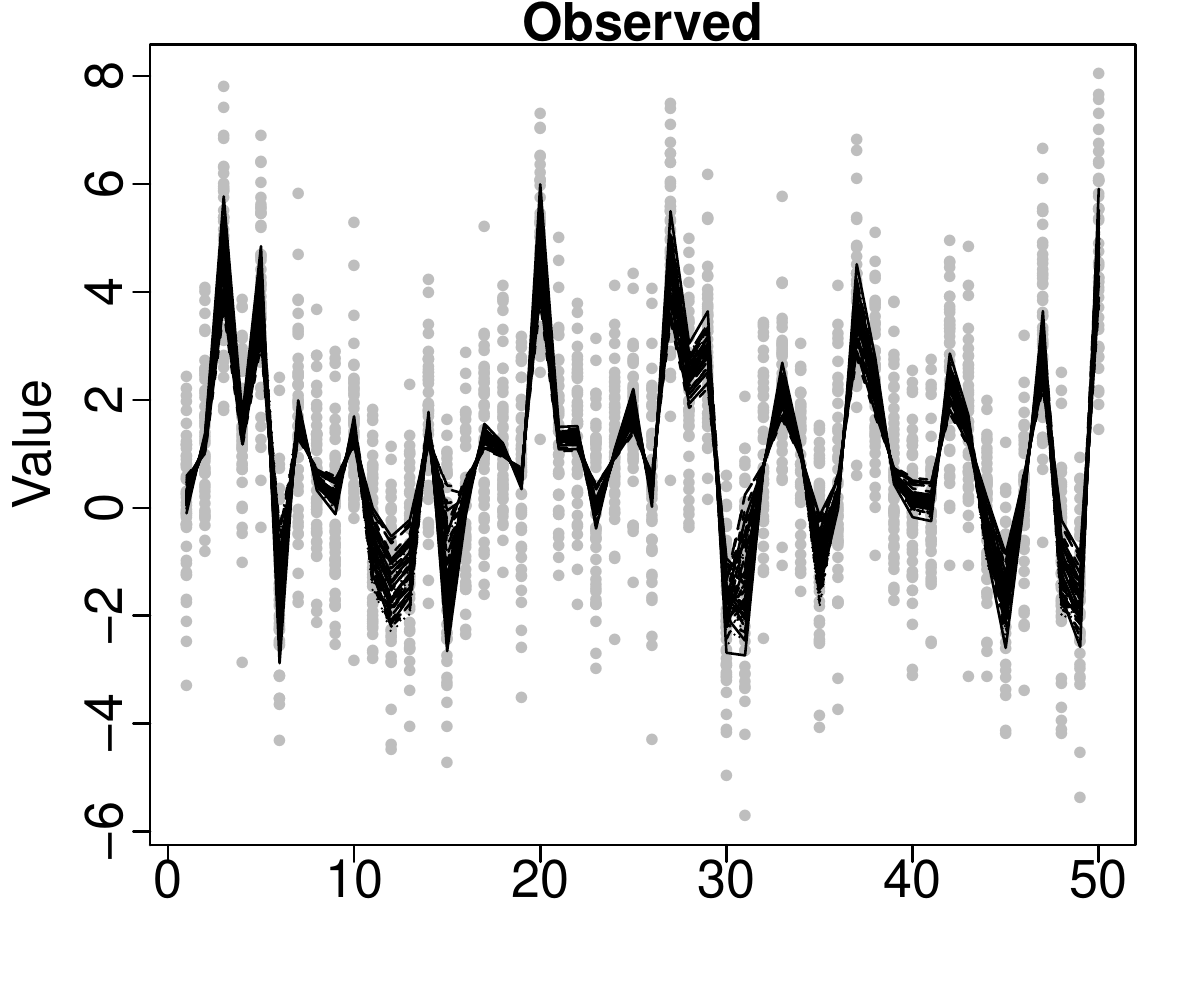}
  \includegraphics[width=5.9cm]{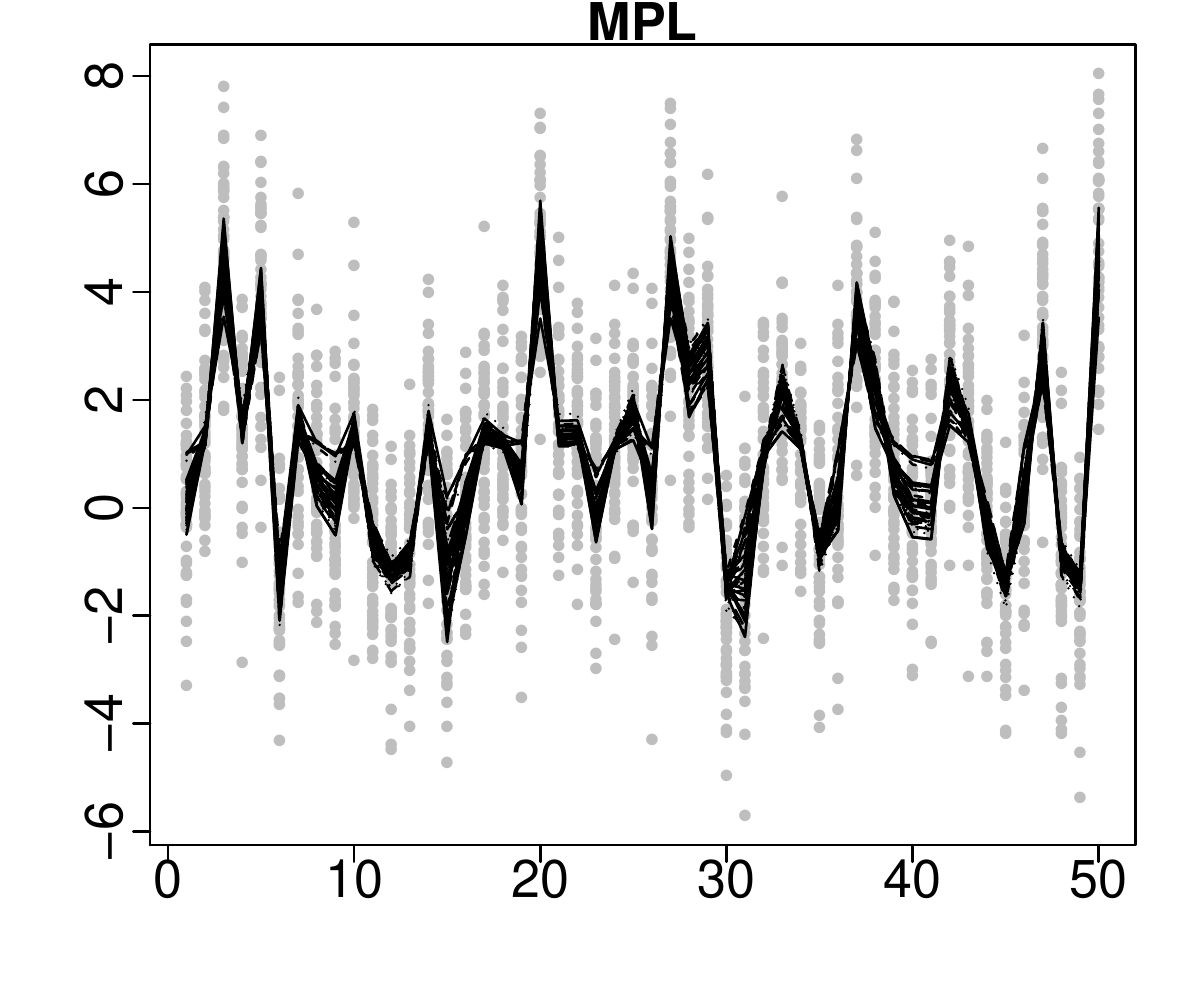}
  \includegraphics[width=5.9cm]{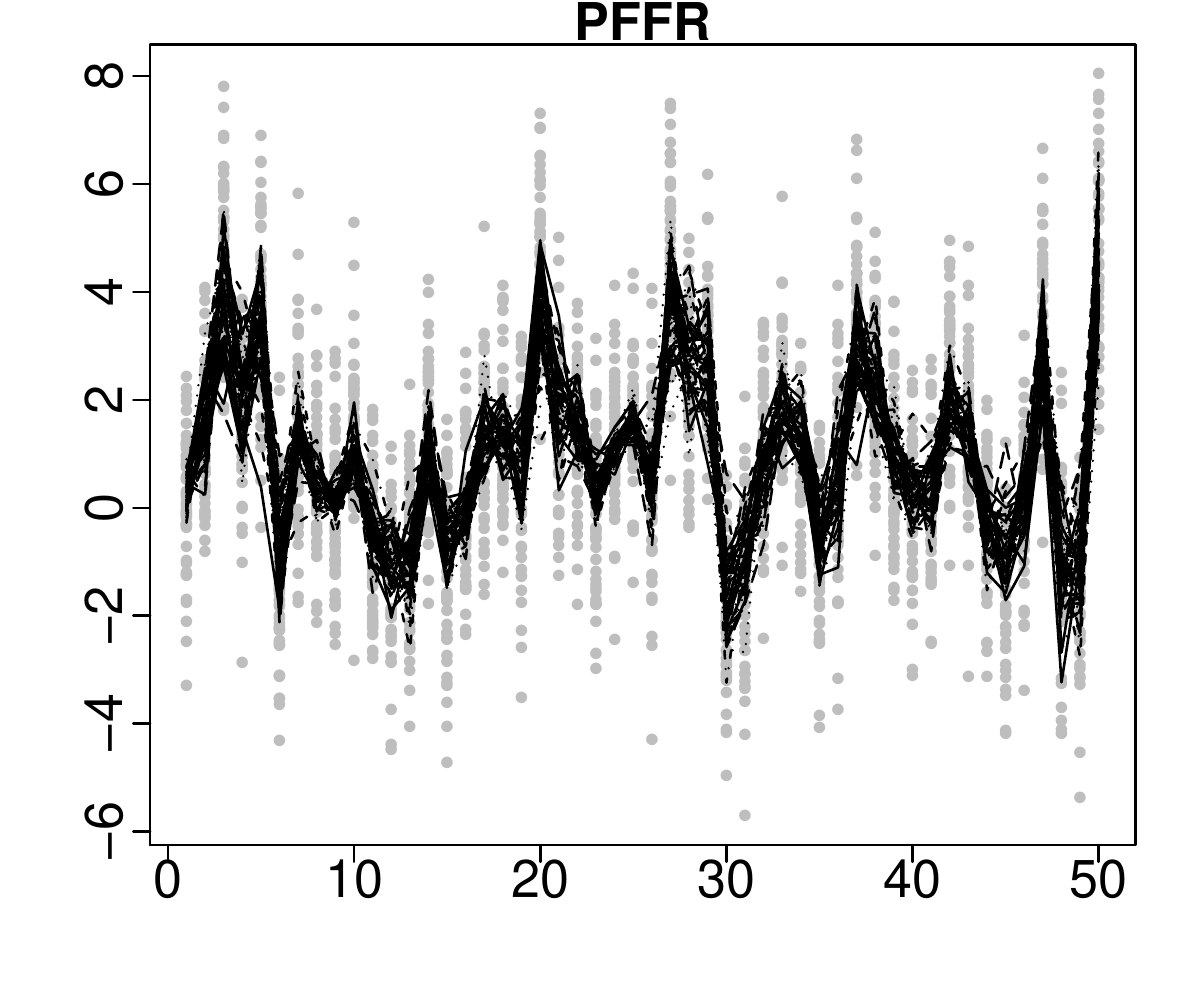}
\\  
  \includegraphics[width=5.9cm]{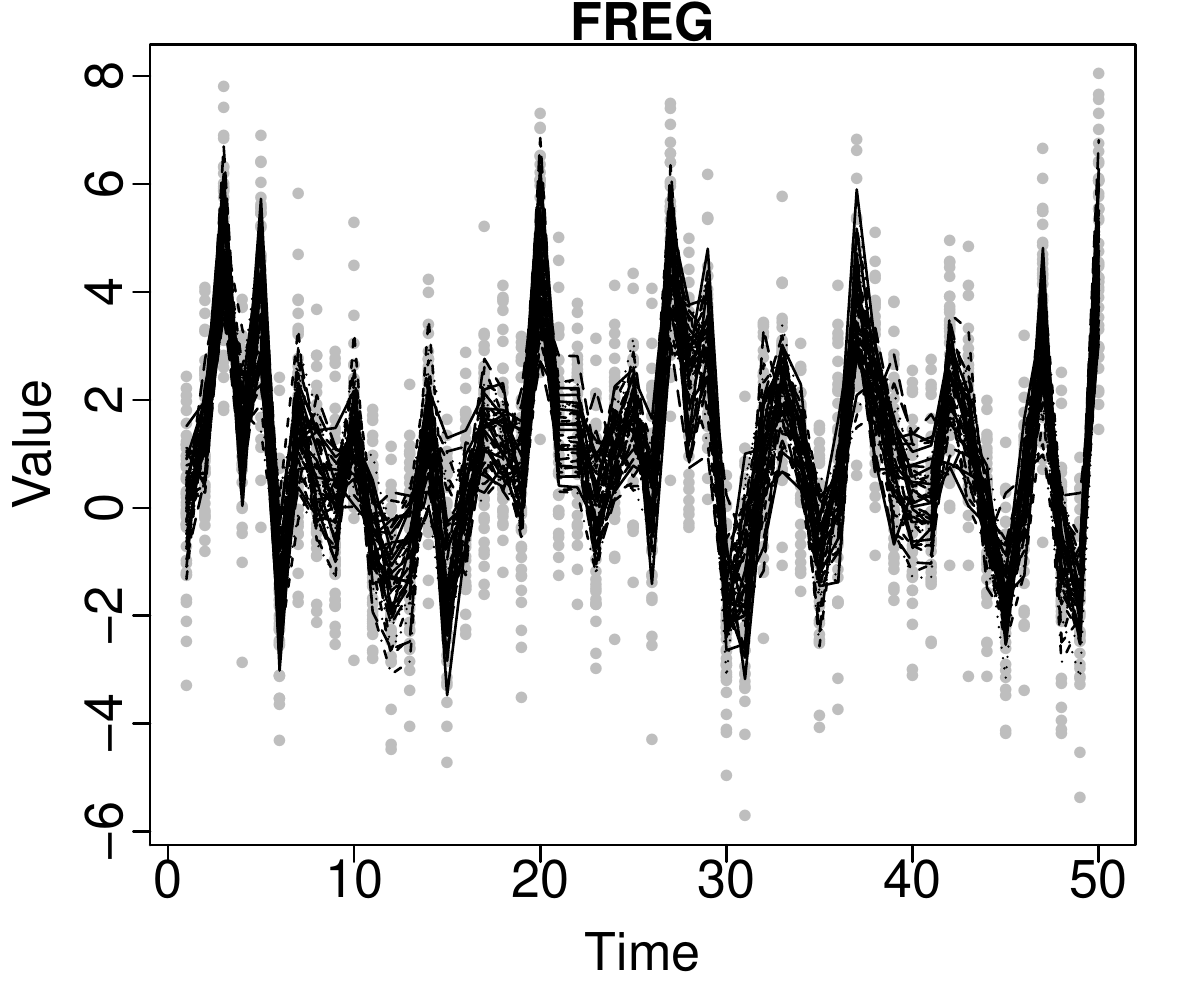}
    \includegraphics[width=5.9cm]{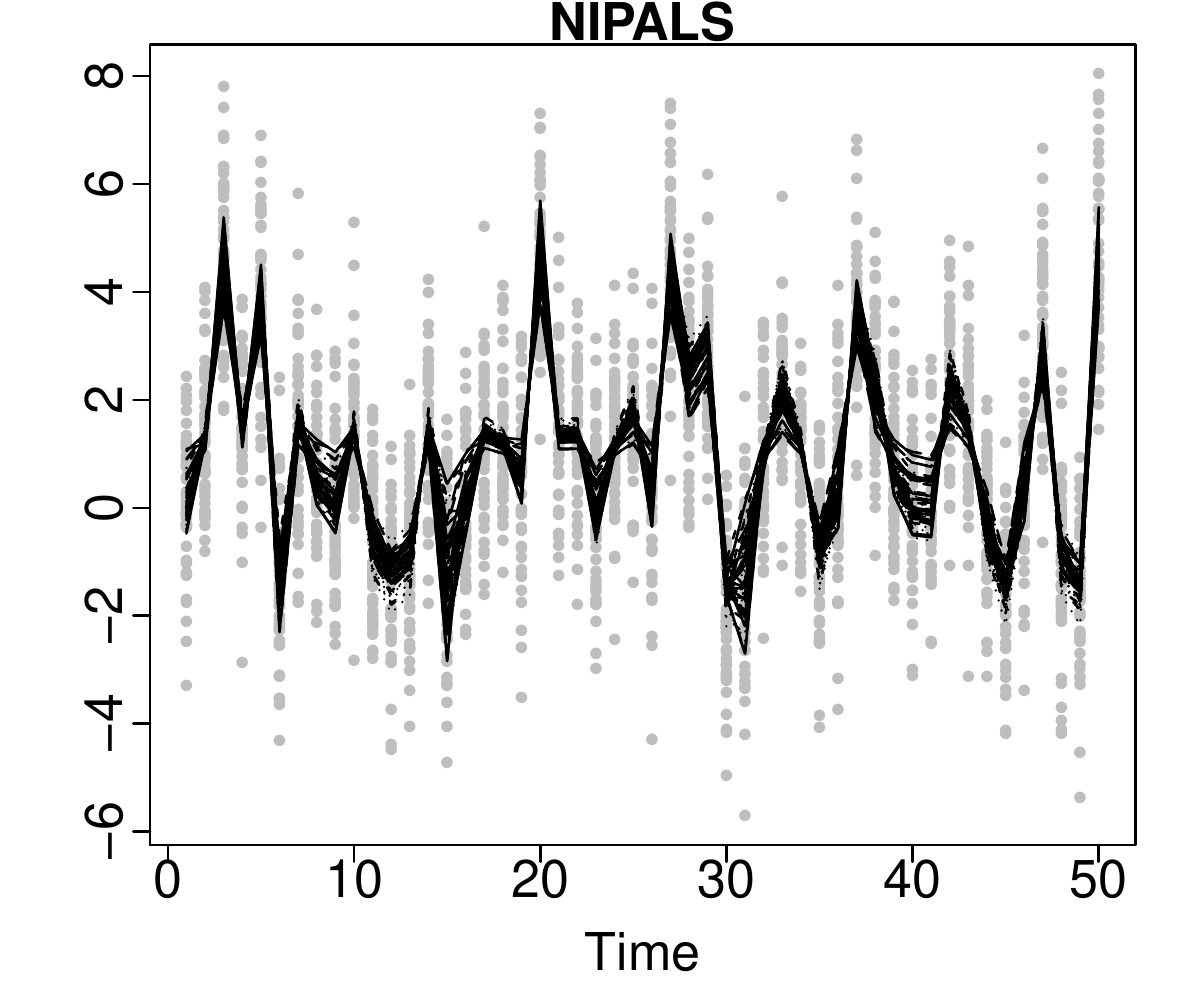}
  \includegraphics[width=5.9cm]{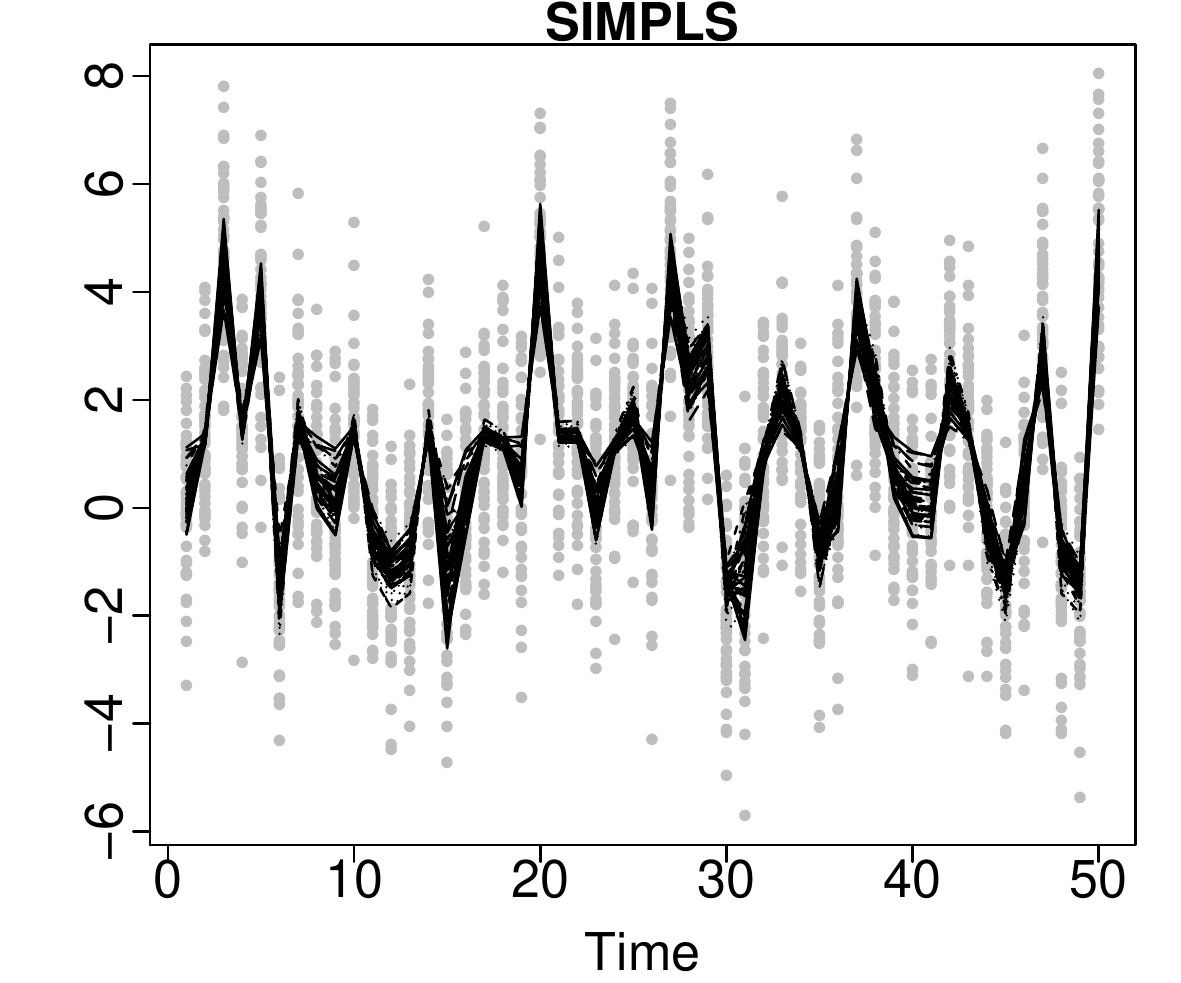}
  \caption{Plots of the generated $N$ sets of discrete data (gray points) and fitted smooth functions (black lines) when $\rho = 1$ and $K = 20$ numbers of basis functions were used in $B$-spline basis. The MPL, PFFR, FREG, NIPALS, and SIMPLS were used to obtain fitted response functions.}
  \label{fig:Fig_1}
\end{figure}

The data generated at discrete time points were first converted into functions using the $B$-spline basis with $K = [10, 20, 30, 40]$ numbers of basis functions. An example of the observed data with noise and the fitted smooth functions for the generated response variable is presented in Figure~\ref{fig:Fig_1}.

\begin{figure}[htbp]
\centering
\subfloat[$\rho = 0.5$]
{\includegraphics[width=85mm]{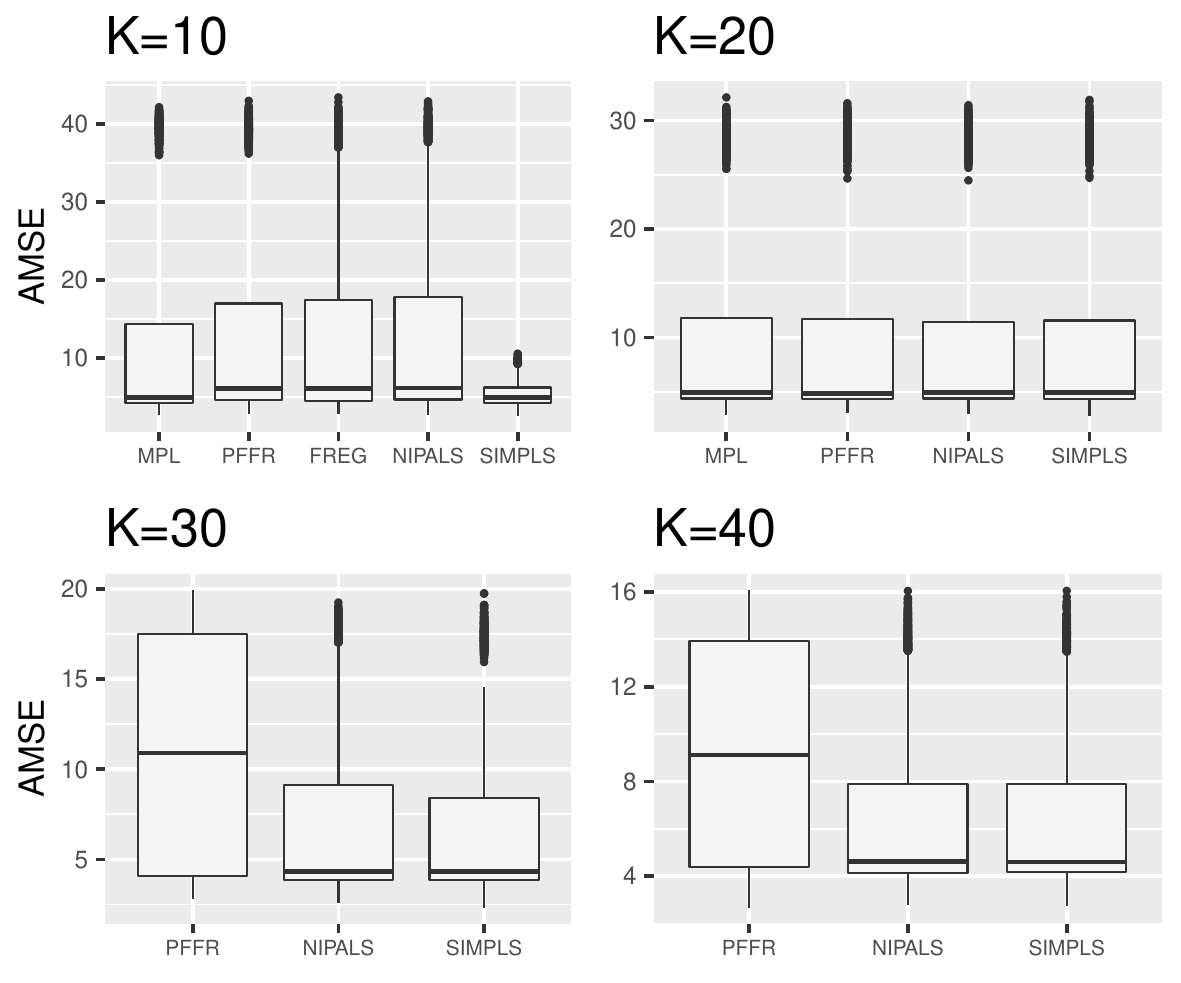}}
\quad
\subfloat[$\rho = 1$]
{\includegraphics[width=85mm]{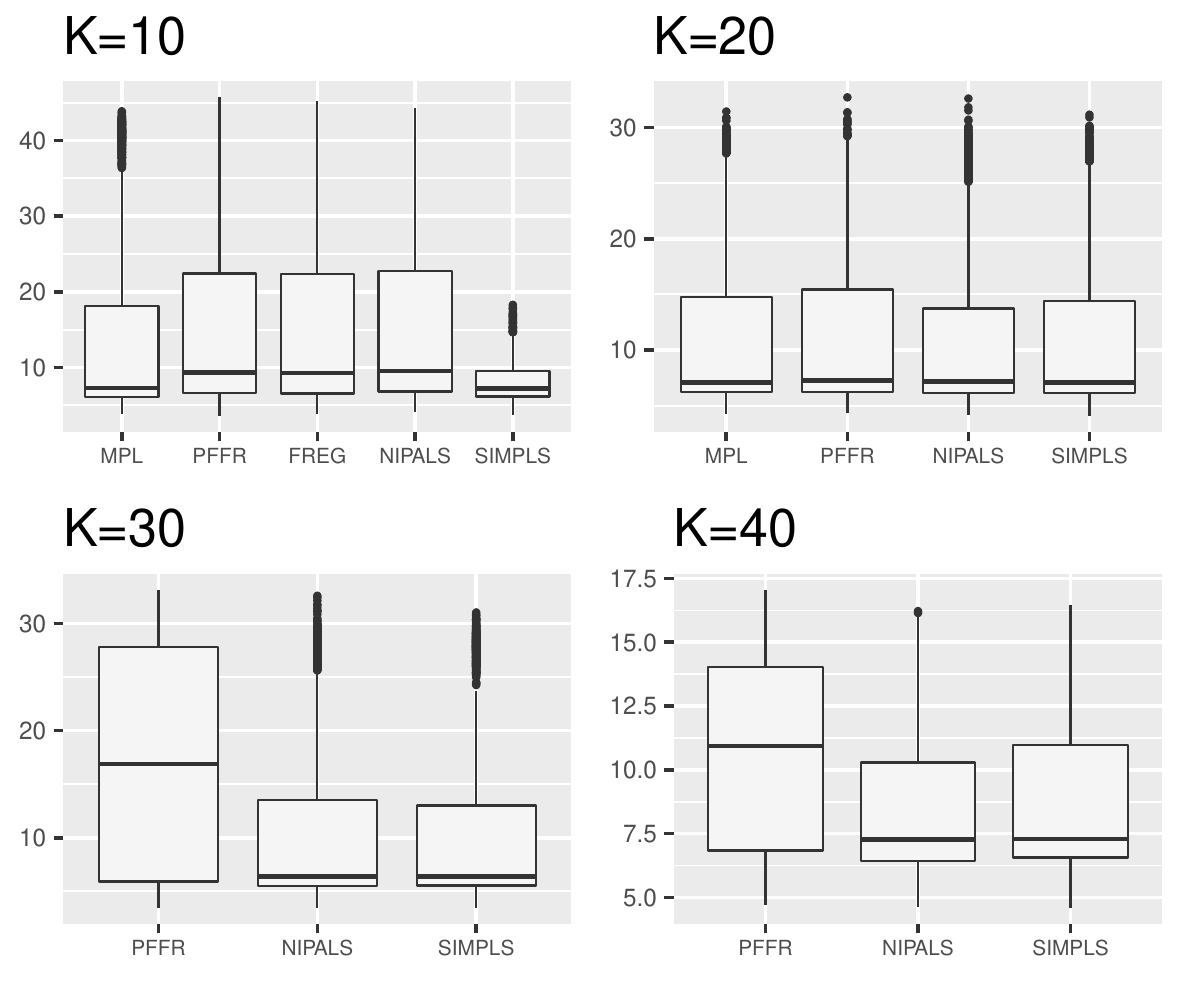}}
\\
\subfloat[$\rho = 2$]
{\includegraphics[width=85mm]{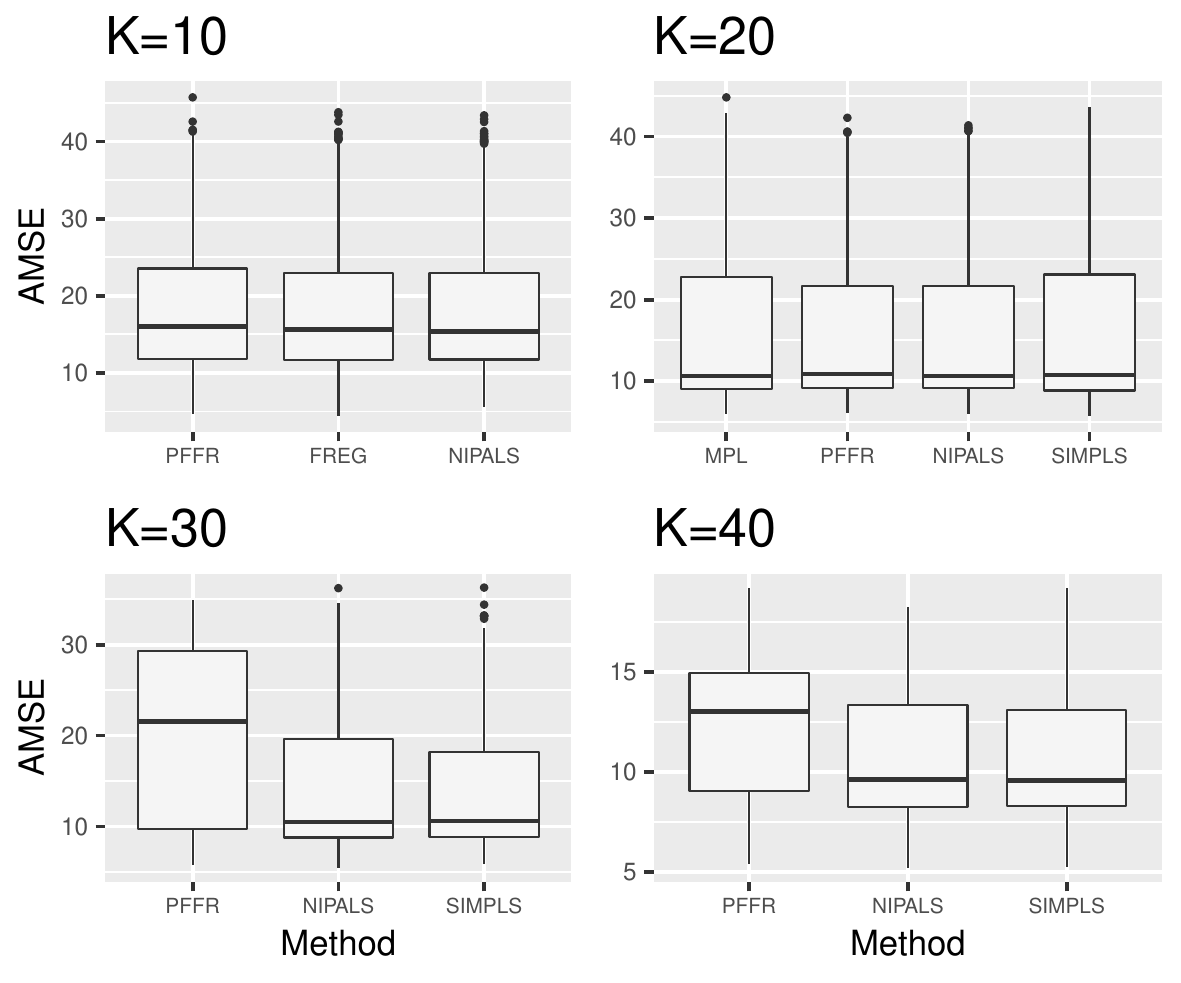}}
\quad
\subfloat[$\rho = 4$]
{\includegraphics[width=85mm]{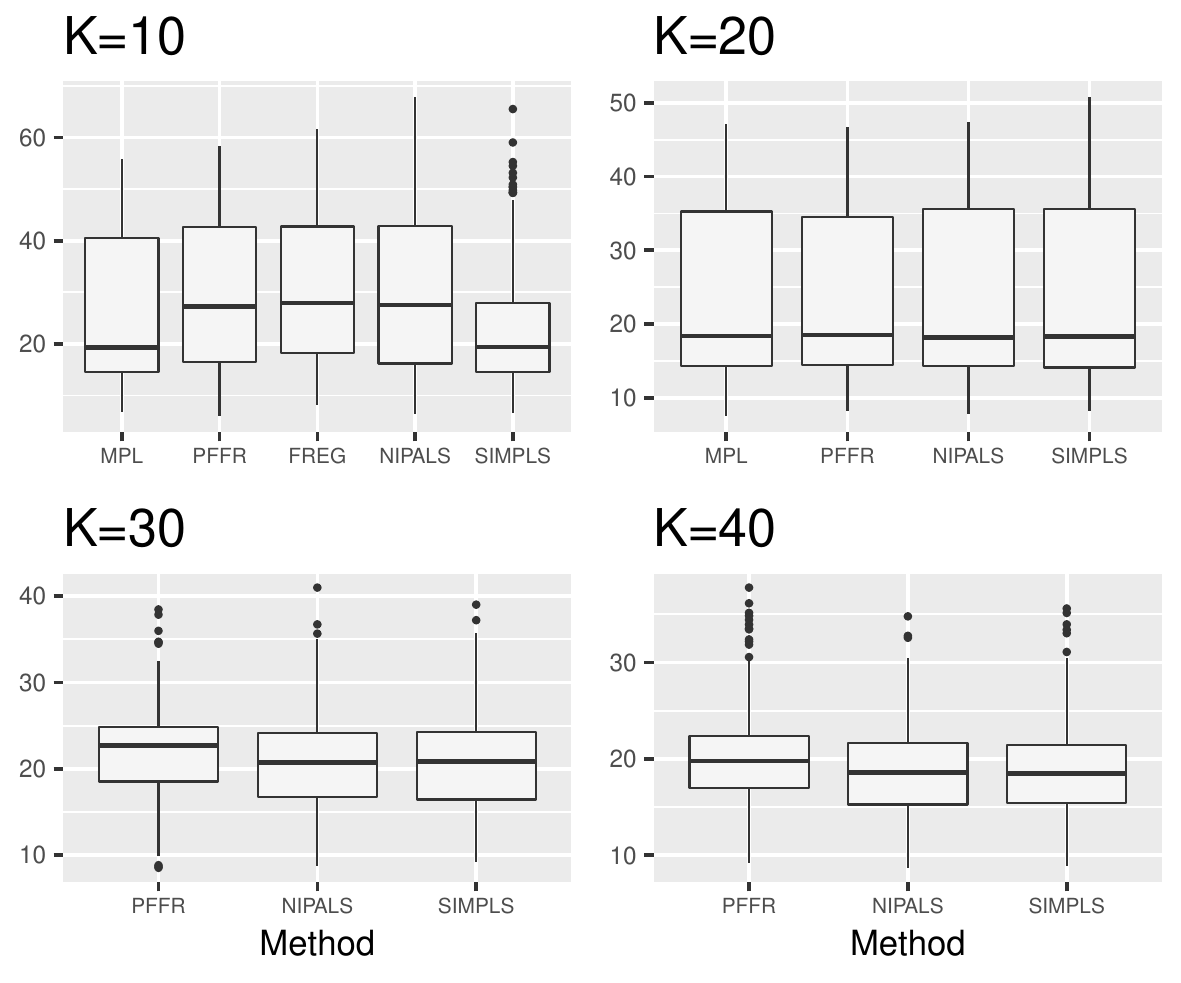}}
\caption{Fitted model performances: Computed AMSE values of the MPL, PFFR, FREG, NIPALS, and SIMPLS methods. The data were generated based on the variance parameter $\rho = [0.5, 1, 2, 4]$ and $K = [10, 20, 30, 40]$ numbers of basis functions were used to convert the data to smooth functions.}
\label{fig:Fig_2}
\end{figure}

\begin{figure}[htbp]
\centering
\subfloat[$\rho = 0.5$]
{\includegraphics[width=88mm]{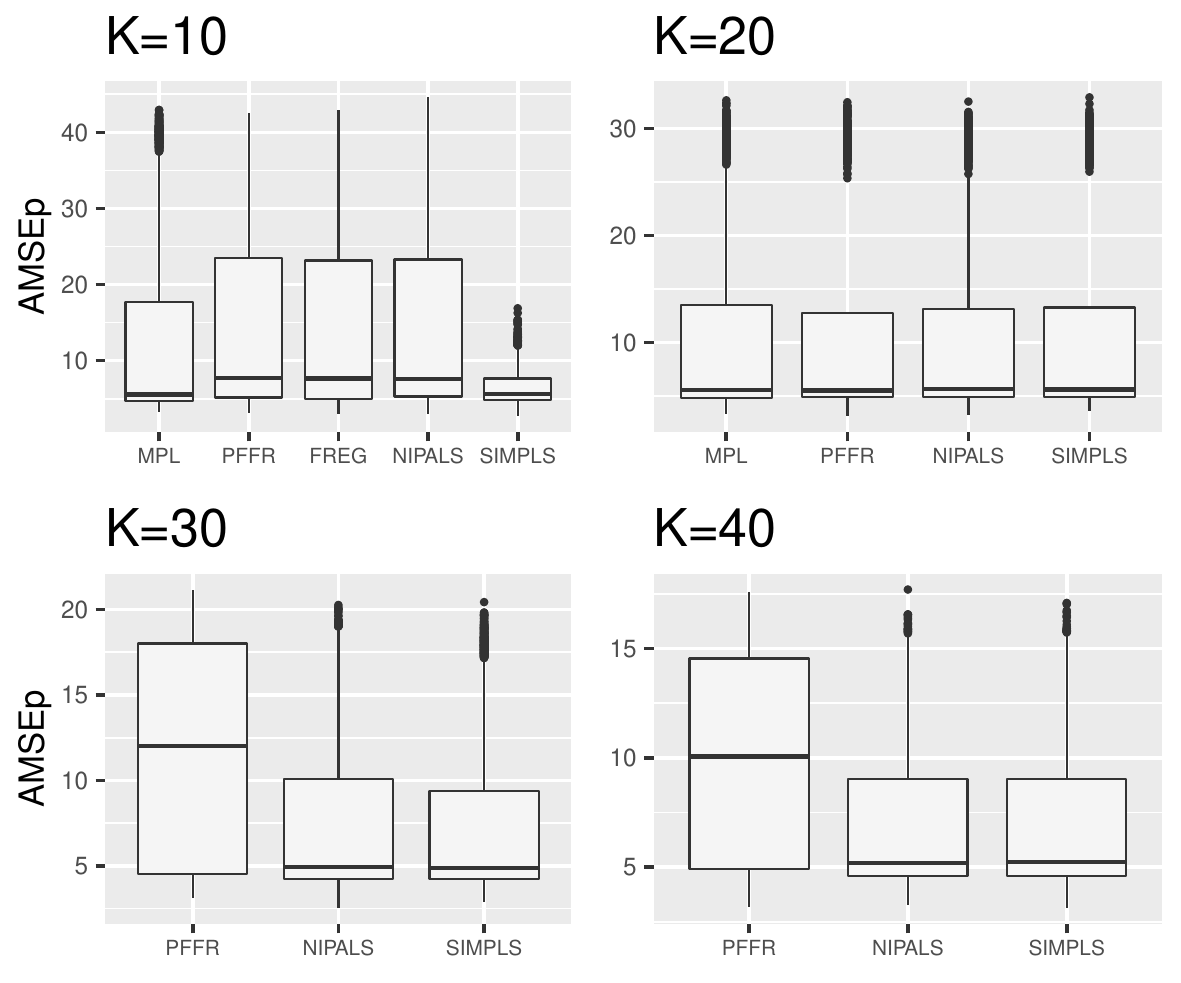}}
\quad
\subfloat[$\rho = 1$]
{\includegraphics[width=88mm]{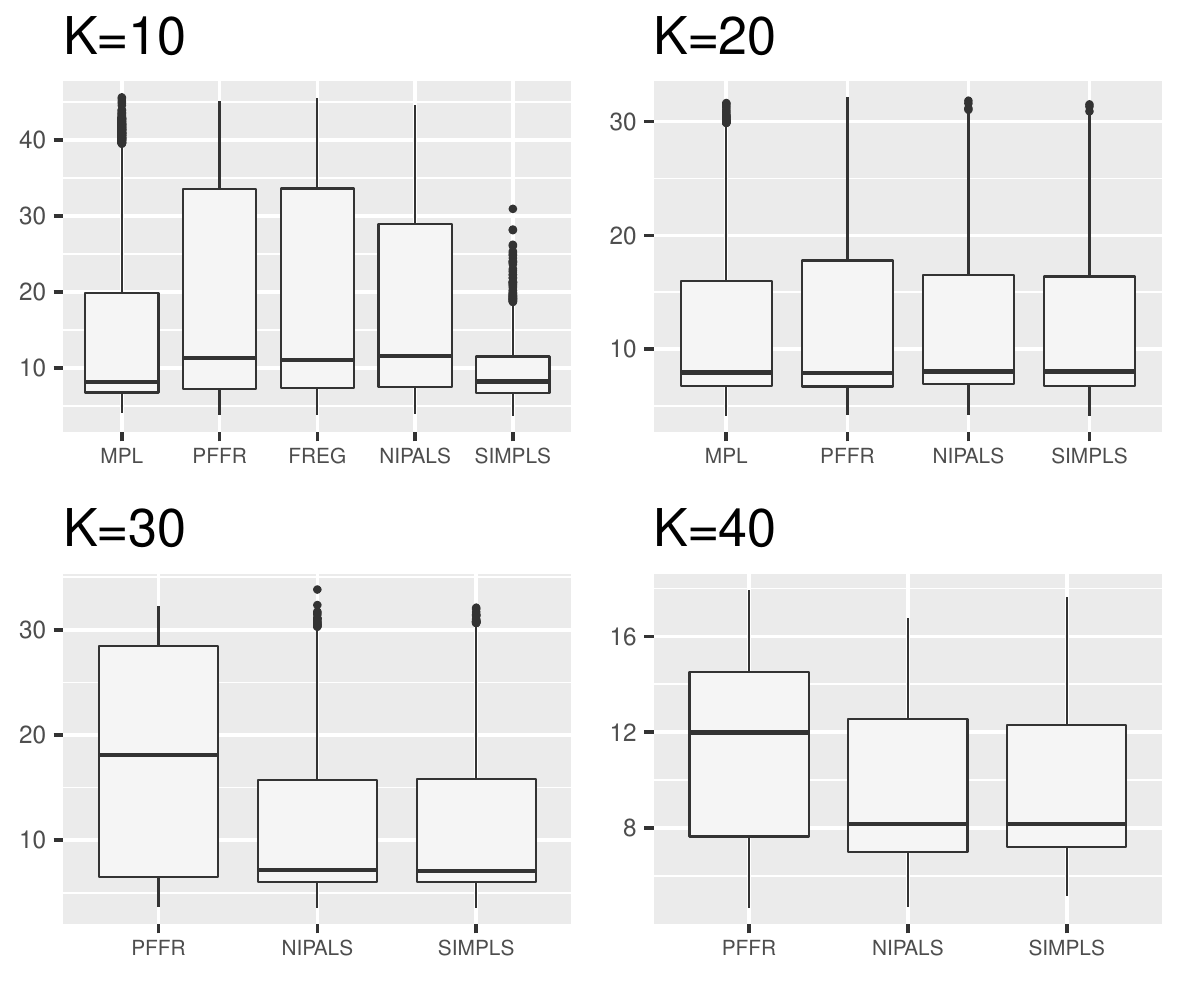}}
\\
\subfloat[$\rho = 2$]
{\includegraphics[width=88mm]{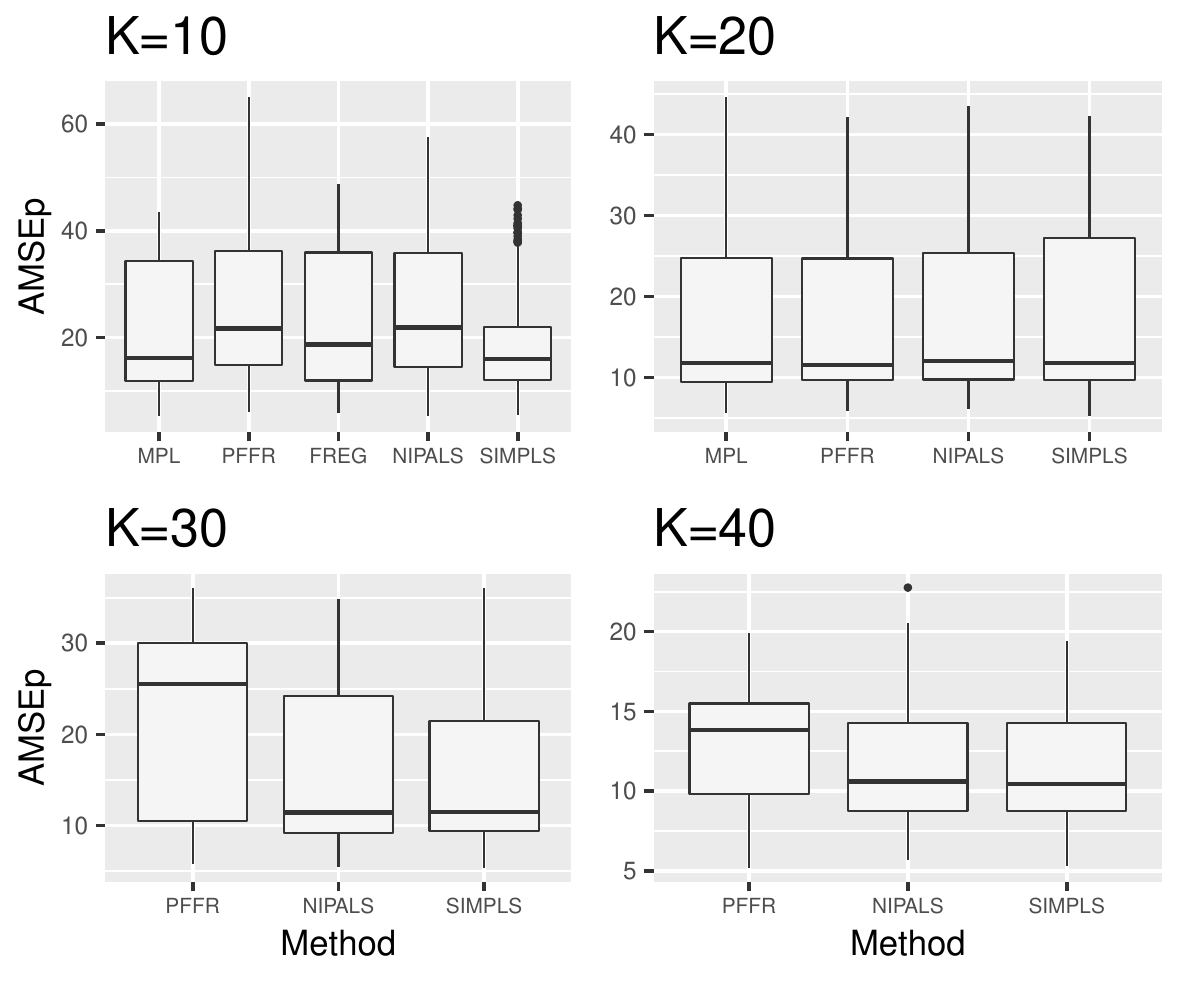}}
\quad
\subfloat[$\rho = 4$]
{\includegraphics[width=88mm]{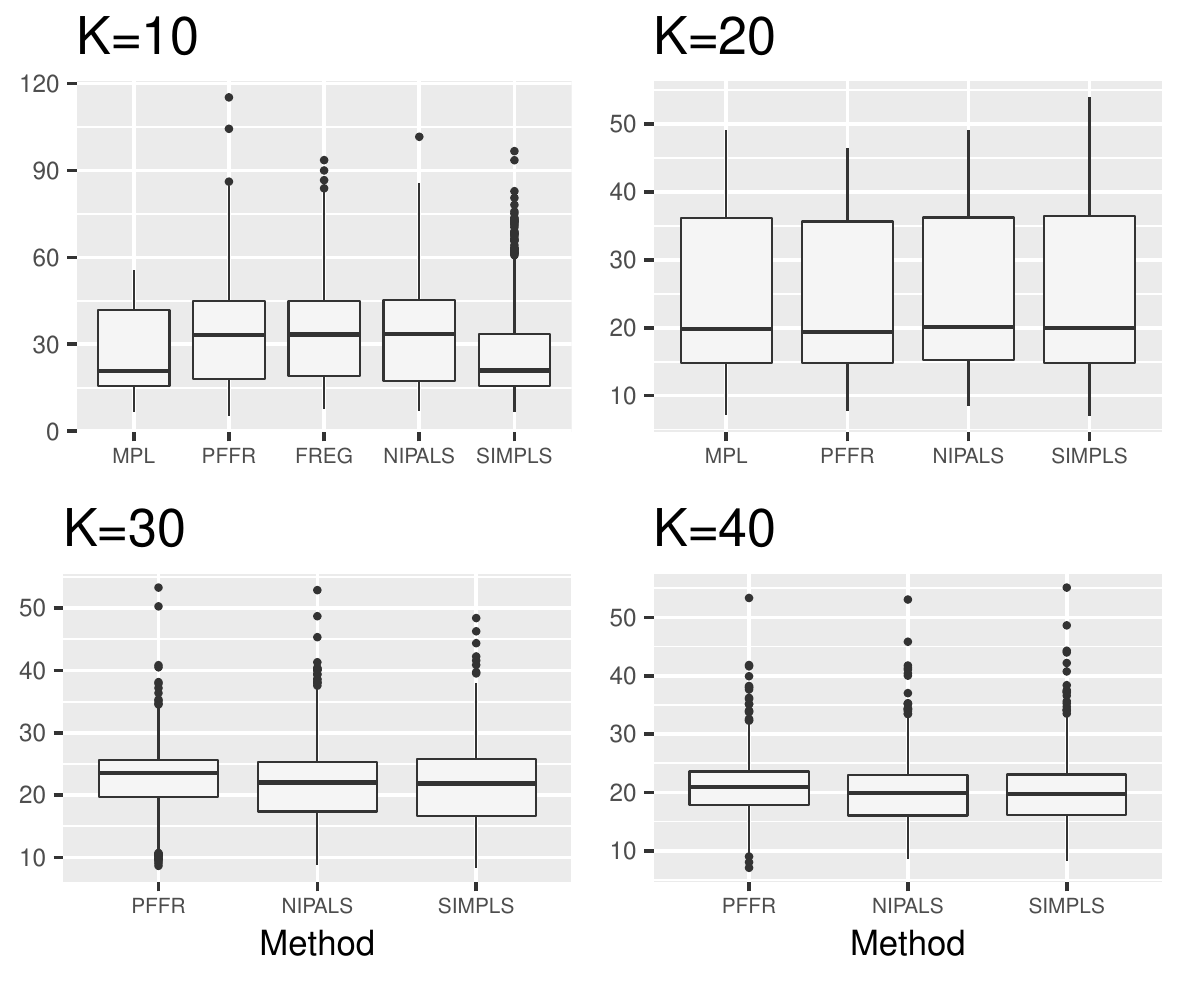}}
\caption{Predicted model performances: Computed AMSE$_p$ values of the MPL, PFFR, FREG, NIPALS, and SIMPLS methods. Data were generated based on the variance parameter $\rho = [0.5, 1, 2, 4]$ and $K = [10, 20, 30, 40]$ numbers of basis functions were used to convert the data to smooth functions.}
\label{fig:Fig_3}
\end{figure}

Before presenting our findings, we note that the results do not vary considerably with different choices of $N$; therefore, to save space, we only report the results for $N = 100$. The LS and ML methods failed to provide an estimate for $\mathbf{B}$ because of the singular matrix problem and degenerate structure of the generated data; thus, we only report on the comparative studies with MPL, PFFR, FREG, NIPALS, and SIMPLS. Our results obtained from the fitted and predicted models are presented in Figures~\ref{fig:Fig_2} and~\ref{fig:Fig_3}, respectively. 

They illustrate that, when $K = 10$, the proposed SIMPLS algorithm performed considerably better than the other methods in terms of AMSE, $\text{AMSE}_p$, and their associated standard errors. Also, the NIPALS algorithm showed competitive performance to other methods. We observed that the FREG and MPL failed to provide an estimate for the model parameter when $K \geq 20$ and $K \geq 30$, respectively. For a small to moderate variance parameter, the proposed NIPALS and SIMPLS performed better than the PFFR, while all three estimation methods tended to have similar performances when $\rho = 4$.

The results for the MCS analysis are presented in Table~\ref{tab:case1}. The values in this table correspond to the percentages of the superiorities of the methods from 1,000 Monte-Carlo simulations. Our findings demonstrate that the proposed NIPALS and SIMPLS algorithms produced significantly better prediction performances compared with their competitors except when $K = 20$.
\begin{center}
\tabcolsep 0.44in
\begin{small}
\begin{longtable}{@{}llllrr@{}} 
\caption{MCS analysis results.}\\
\toprule
$\rho$ & {Method} & {$K=10$} & {$K=20$} & {$K=30$} & {$K=40$} \\
\midrule
0.5 & MPL	& 0.090\%	& 0.532\%	& -		 	& -		 	\\
 & PFFR		& 0.000\%	& 0.000\%	& 0.000\% 	& 0.000\%  \\
 & FREG		& 0.000\% 	& -		 	& -		 	& -		 	\\
 & NIPALS	& 0.610\% 	& 0.299\% 	& 0.547\% 	& 0.588\%  \\
 & SIMPLS	& 0.318\% 	& 0.197\% 	& 0.467\% 	& 0.424\%  \\
\midrule
1 & MPL		& 0.183\% 	& 0.603\% 	& -		 	& -		 	\\
 & PFFR		& 0.000\% 	& 0.000\% 	& 0.000\% 	& 0.000\%  \\
 & FREG		& 0.000\% 	& -		 	& -		 	& -		 	\\
 & NIPALS	& 0.529\% 	& 0.257\% 	& 0.485\% 	& 0.655\%  \\
 & SIMPLS	& 0.301\% 	& 0.161\% 	& 0.528\% 	& 0.352\%  \\
\midrule
2 & MPL		& 0.001\% 	& 0.517\% 	& -		 	& -		 	\\
 & PFFR		& 0.000\% 	& 0.000\% 	& 0.000\% 	& 0.004\%  \\
 & FREG		& 0.000\% 	& -		 	& -		 	& -		 	\\
 & NIPALS	& 0.321\% 	& 0.143\% 	& 0.478\% 	& 0.533\%  \\
 & SIMPLS	& 0.682\% 	& 0.354\% 	& 0.526\% 	& 0.447\%  \\
\midrule
4 & MPL		& 0.644\% 	& 0.803\% 	& -		 	& -		 	\\
 & PFFR		& 0.000\% 	& 0.000\% 	& 0.147\% 	& 0.163\%  \\
 & FREG		& 0.000\% 	& -		 	& -		 	& -		 	\\
 & NIPALS	& 0.212\% 	& 0.192\% 	& 0.410\% 	& 0.494\%  \\
 & SIMPLS	& 0.150\% 	& 0.008\% 	& 0.450\% 	& 0.357\%  \\
\bottomrule
\label{tab:case1}
\end{longtable}
\end{small}
\end{center}

\begin{figure}[!htbp]
  \centering
  \includegraphics[width=5.8cm]{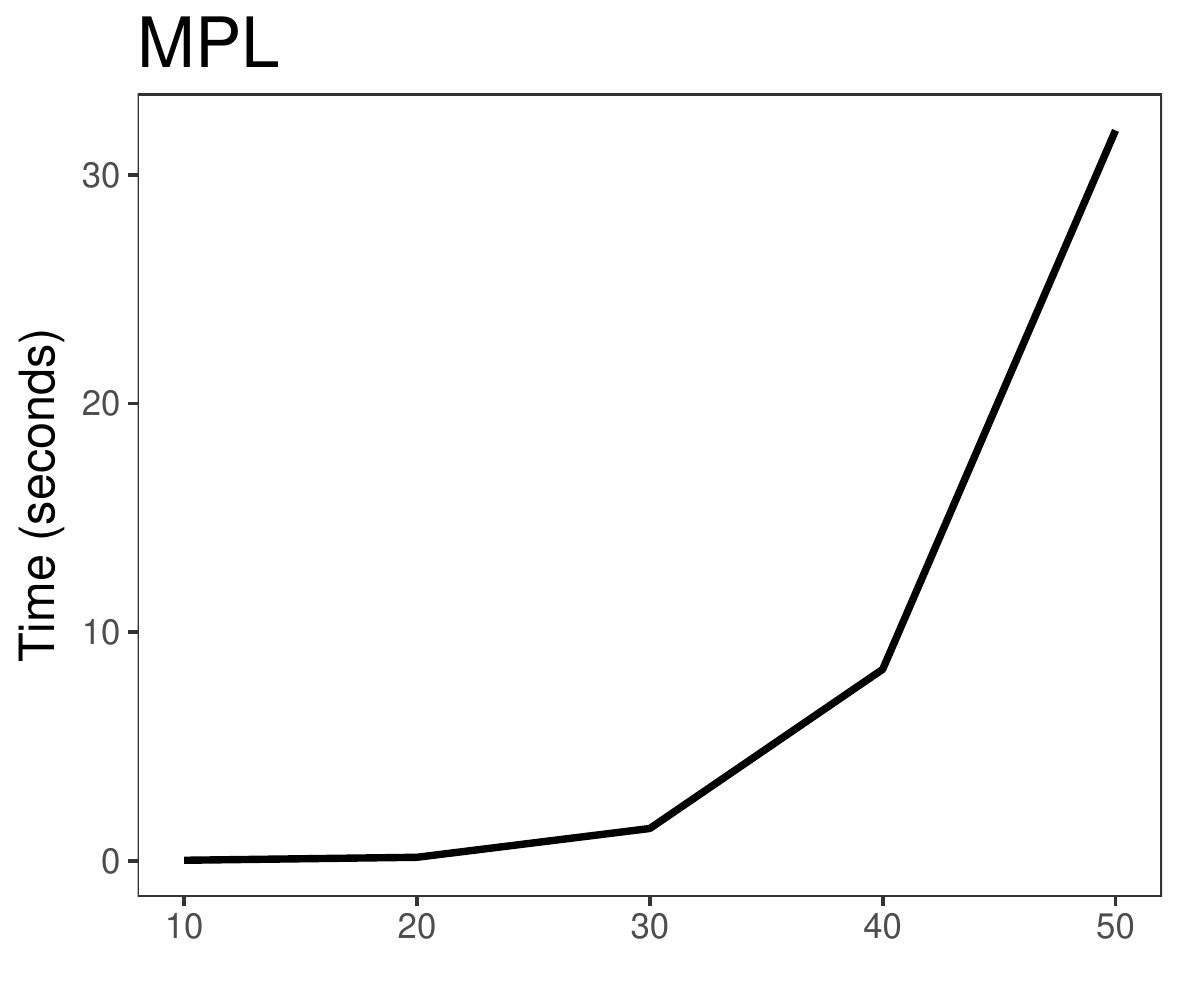}
  \includegraphics[width=5.8cm]{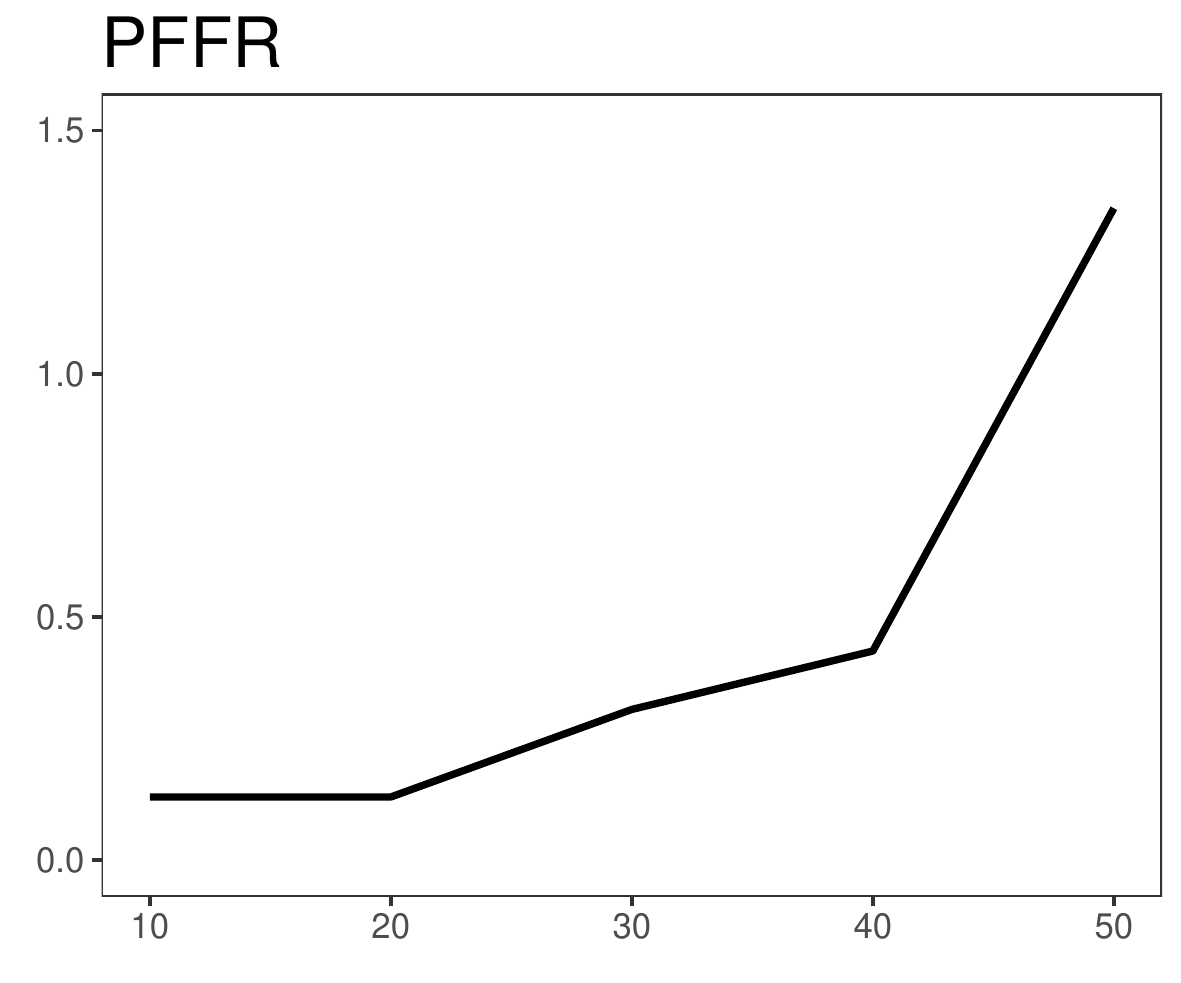}
  \includegraphics[width=5.8cm]{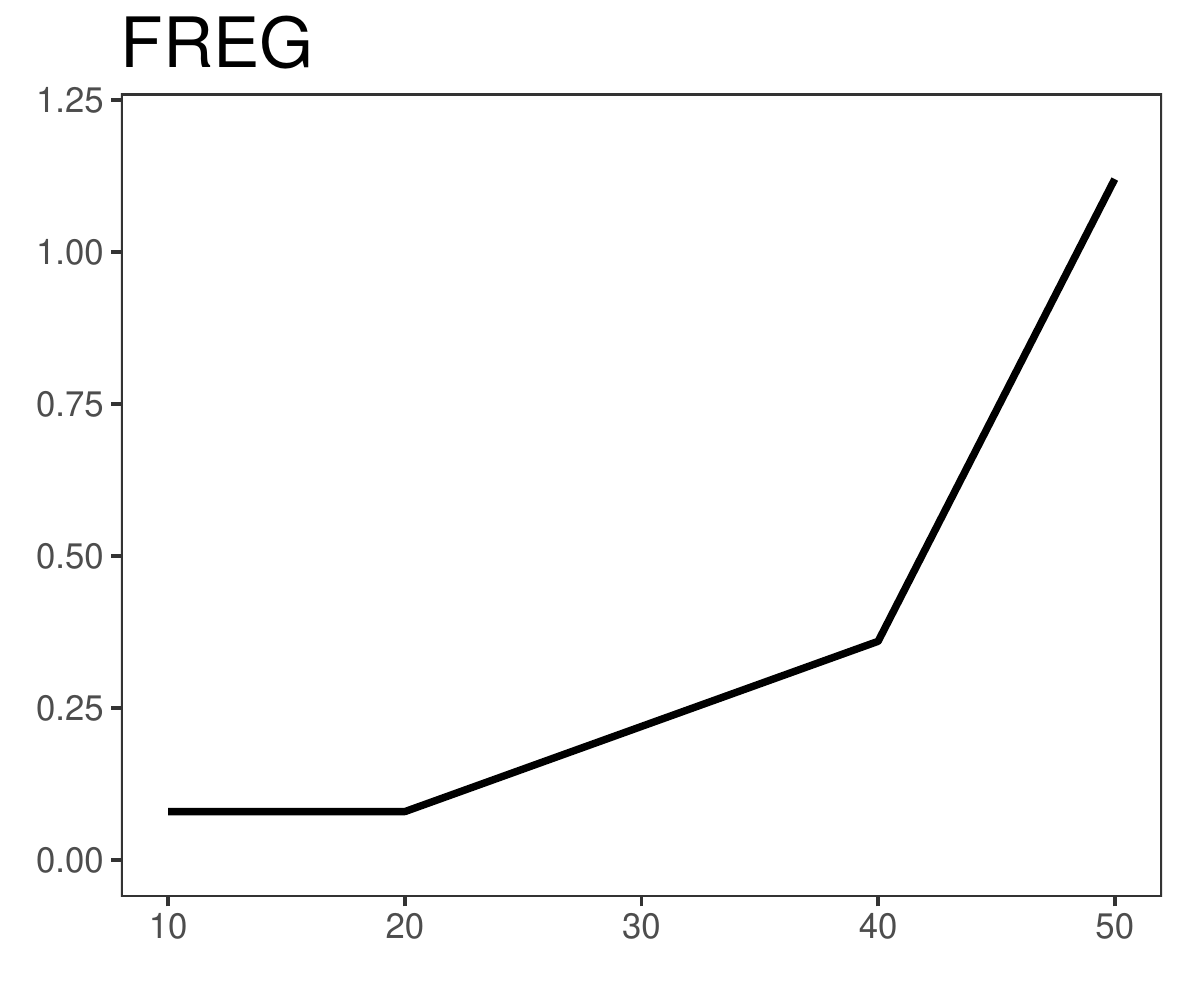}
\\
  \includegraphics[width=5.8cm]{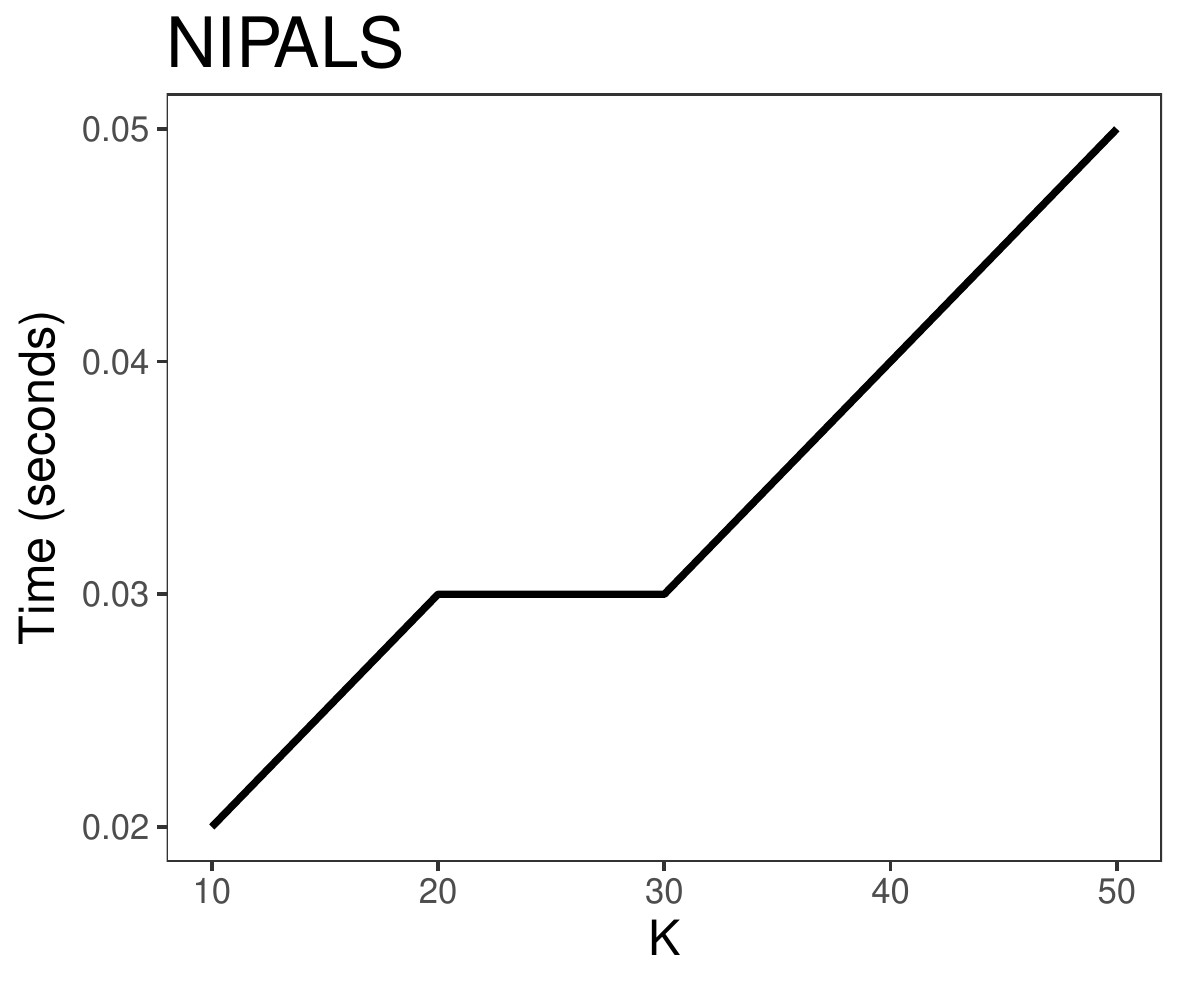}
  \includegraphics[width=5.8cm]{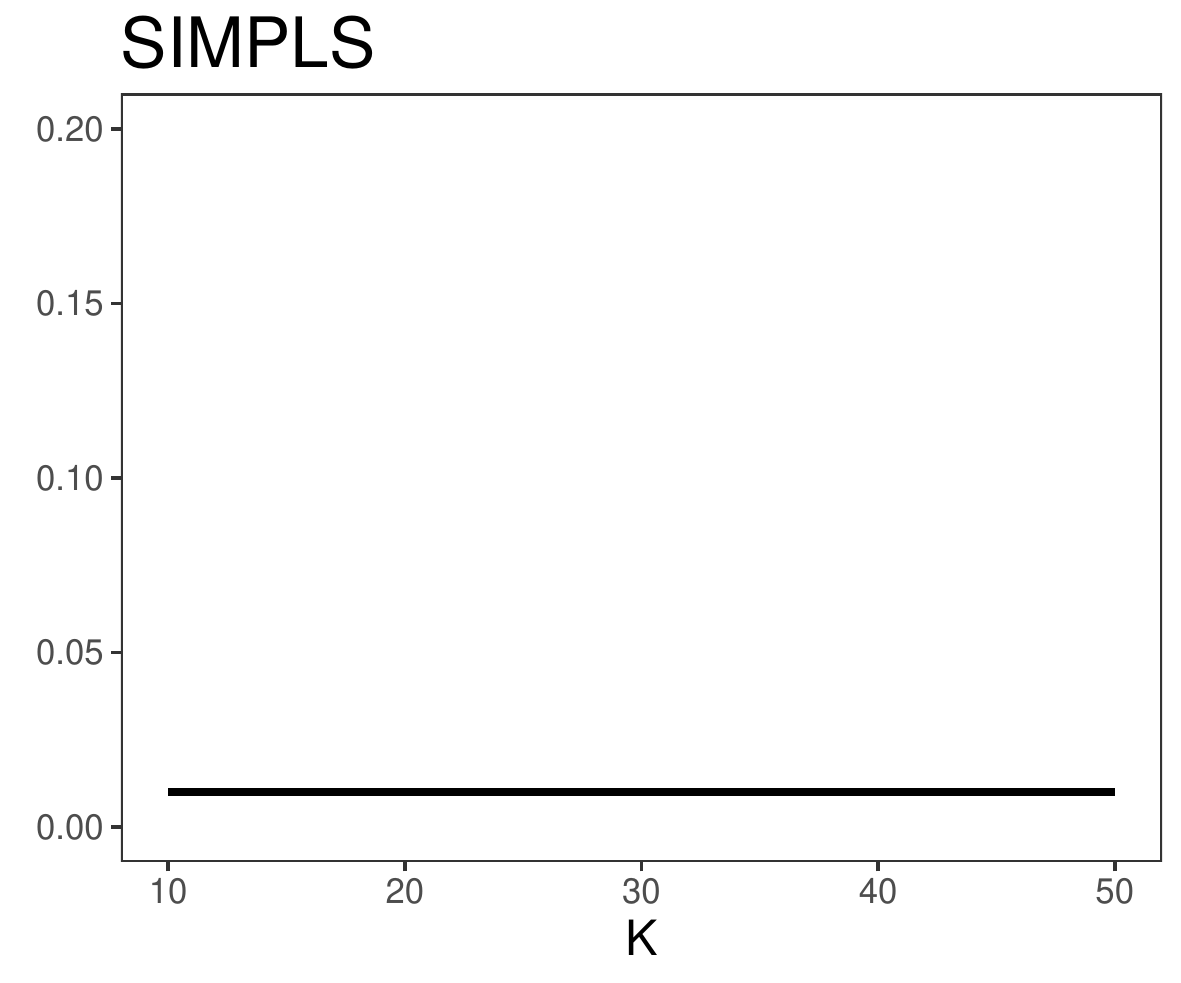}  
  \caption{Estimated computational times in second for the MPL, PFFR, FREG, NIPALS, and SIMPLS estimators.}
  \label{fig:Fig_4}
\end{figure}

Furthermore, we examined the computing performances of the methods considered in this study. Figure~\ref{fig:Fig_4} represents the elapsed computational times for a different number of basis functions obtained by a single Monte-Carlo experiment. This figure illustrates that both the NIPALS and SIMPLS algorithms had considerably shorter computational times than other methods. The computational time of MPL increased exponentially with increasing $K$; therefore, we do not recommend its use when a large number of basis functions are used in the FFRM.

\section{Data Analyses\label{sec:real}}

In this section, we evaluate the performances of the proposed PLS-based methods using an empirical data example: daily North Dakota weather data. The daily dataset was collected from 70 stations across North Dakota (see Table~\ref{tab:stations}), from January 2010 to December 2018 (dataset are available from the North Dakota Agricultural Weather Network Center: \url{https://ndawn.ndsu.nodak.edu}). The dataset has three meteorological variables: average temperature ($^\circ$C), average wind speed (m/sec), and total solar radiation (MJ/m$^2$).

\begin{table}[htbp]
\centering
\tabcolsep 0.11in
\caption{Station names for the North Dakota weather data.}
\begin{tabular}{@{}lllllll@{}}
\toprule
Station & Station & Station & Station & Station & Station & Station  \\
\midrule
Ada		& Cavalier		& Fingal		& Hofflund		 & Marion		 & Perley 		& Sidney \\
Baker	& Crary		& Forest River	& Humboldt	& Mavie		& Pillsbury	& Stephen\\			
Beach	& Crosby		& Galesburg	& Inkster		& Mayville		& Plaza		& Streeter \\	
Berthold	& Dazey  		& Grafton		& Jamestown	& McHenry	& Prosper		& Thomas \\
Bottineau	& Dickinson	& Grand Forks	 & Karlsruhe	& Michigan	& Robinson	& Tappen \\
Bowbells	& Dunn		& Greenbush	 & Langdon	& Minot		& Rolla		& Turtle Lake \\
Bowman	& Edgeley 	 & Harvey		& Leonard		& Mohall		& Roseau		& Wahpeton \\	
Brorson	& Ekre		 & Hazen		& Linton		& Mooreton	& Ross		& Warren\\	
Cando	& Eldred		& Hettinger	& Lisbon		& Mott		& Rugby		& Watford City \\	
Carrington & Fargo		& Hillsboro	& Mandan		& Oakes		& Sabin		& Williston \\	
\bottomrule
\end{tabular}
\label{tab:stations}
\end{table}

\begin{figure}[!htbp]
  \centering
  \includegraphics[width=5.8cm]{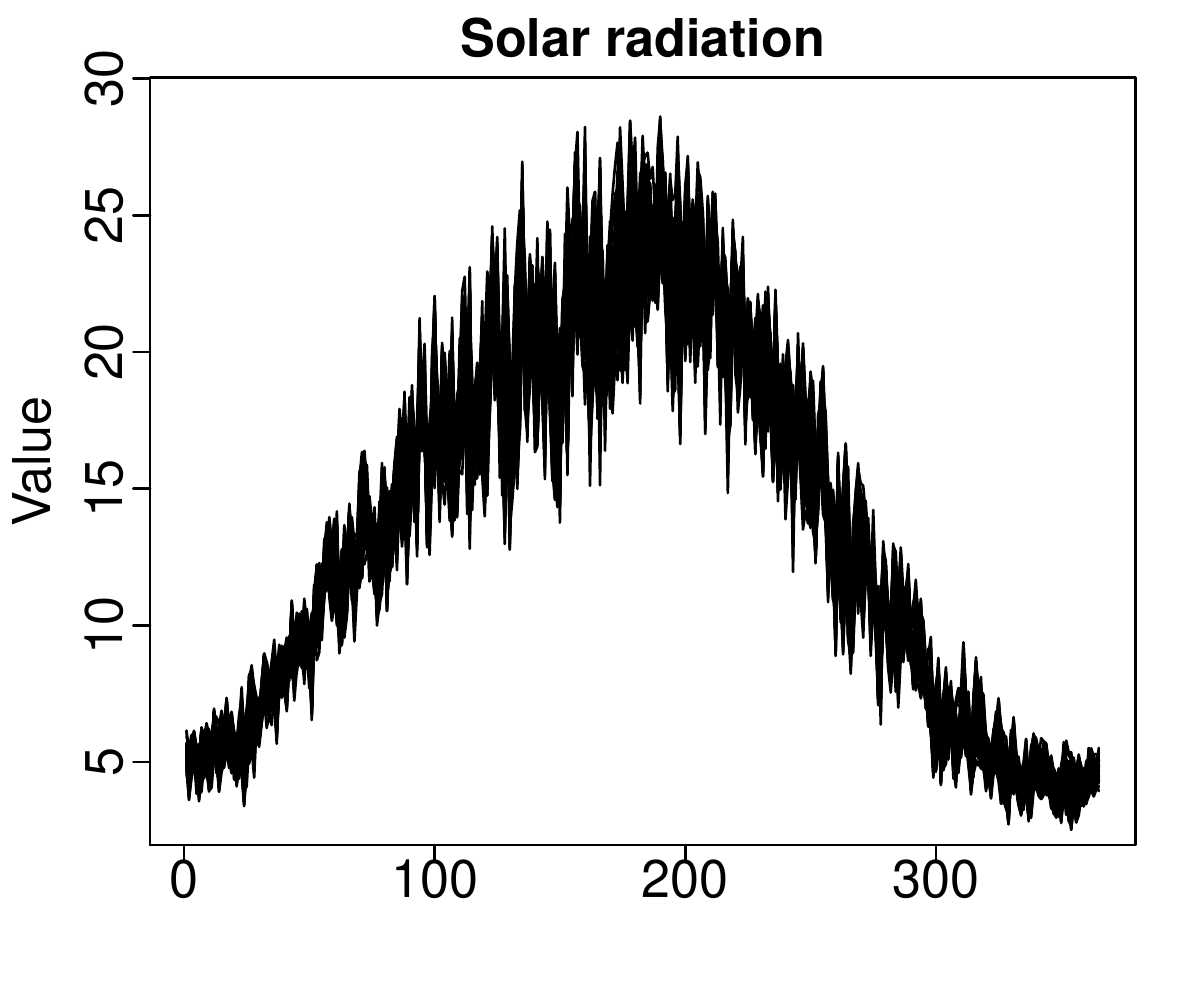}
  \includegraphics[width=5.8cm]{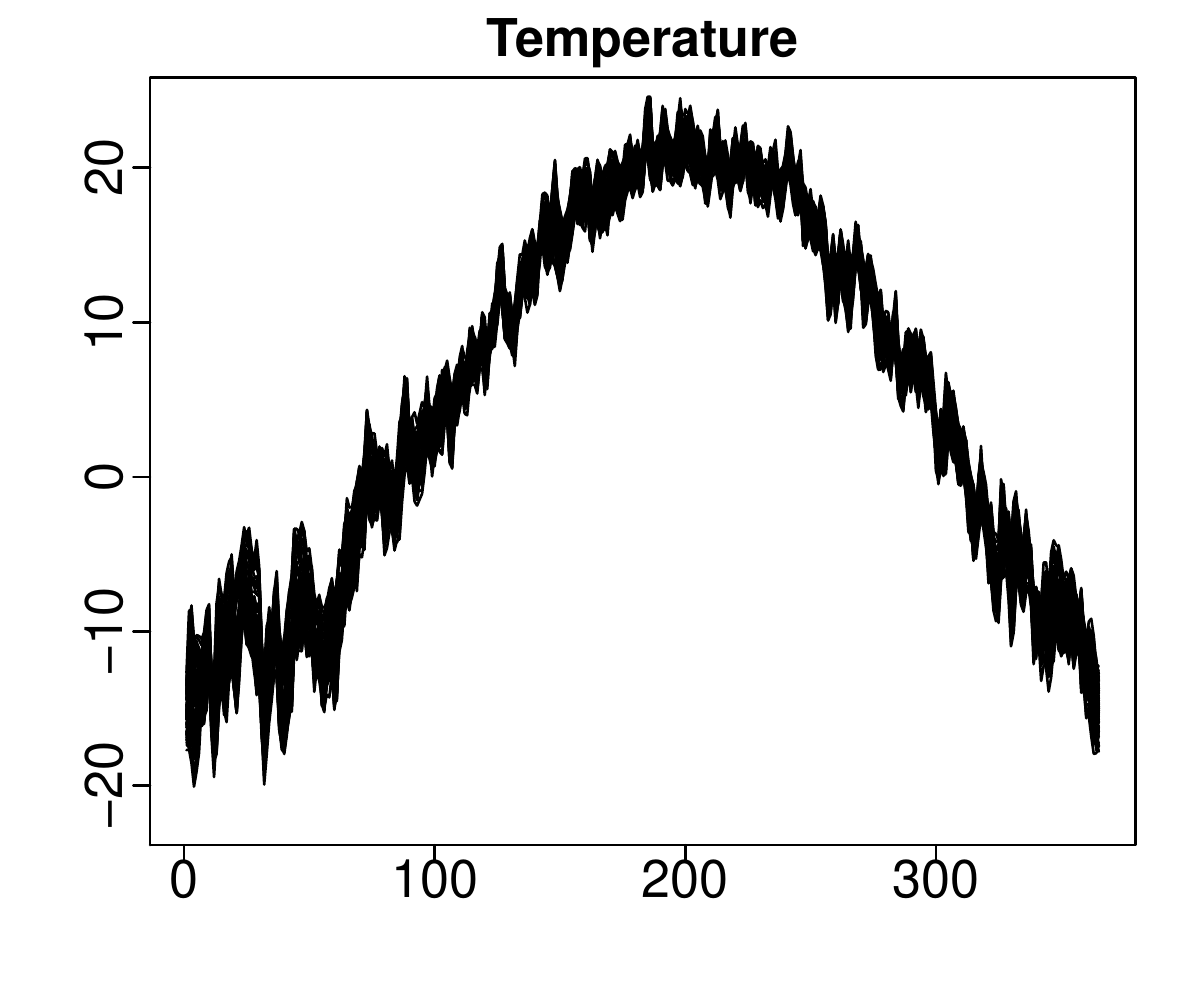}
  \includegraphics[width=5.8cm]{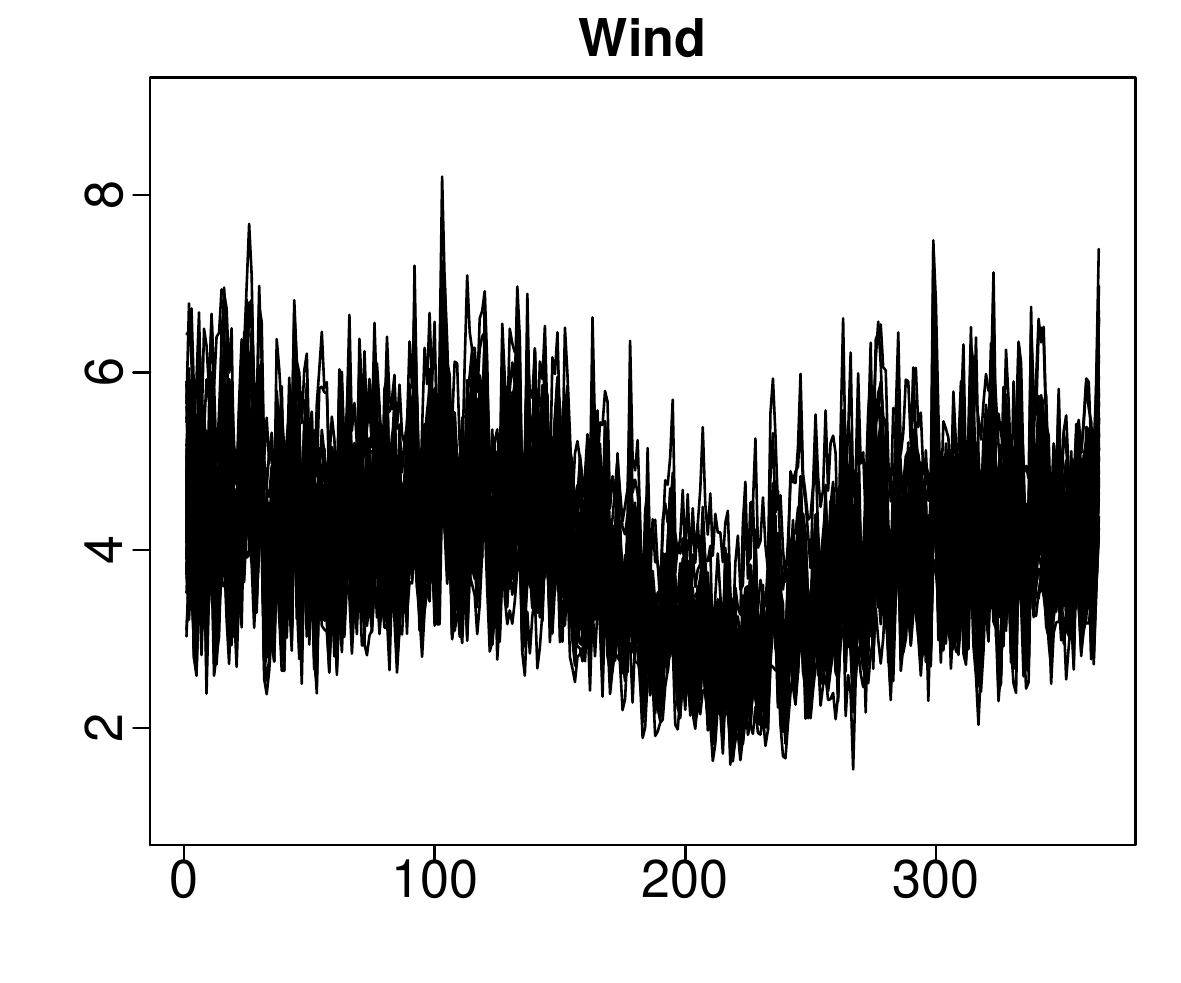}
\\
  \includegraphics[width=5.8cm]{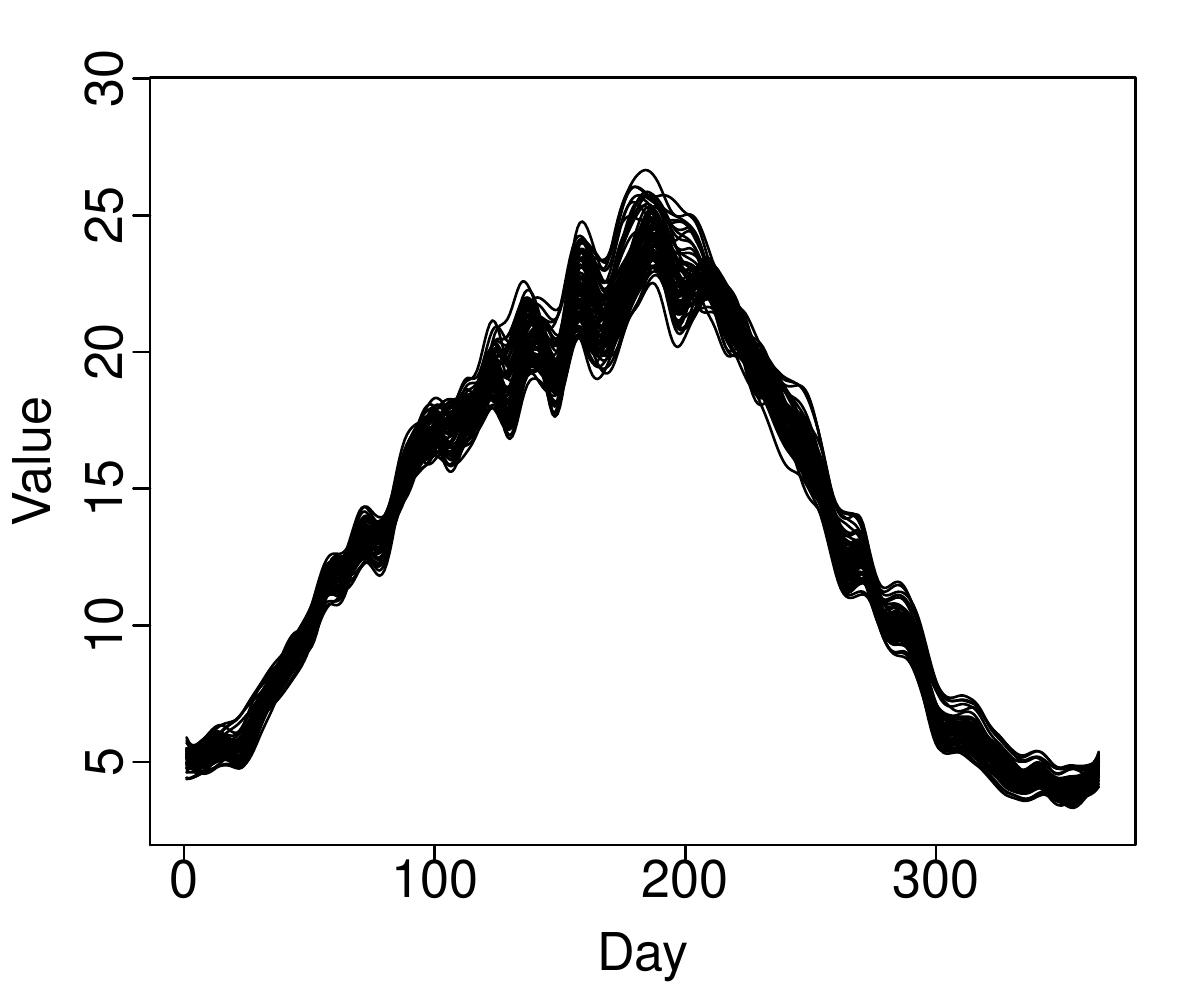}
    \includegraphics[width=5.8cm]{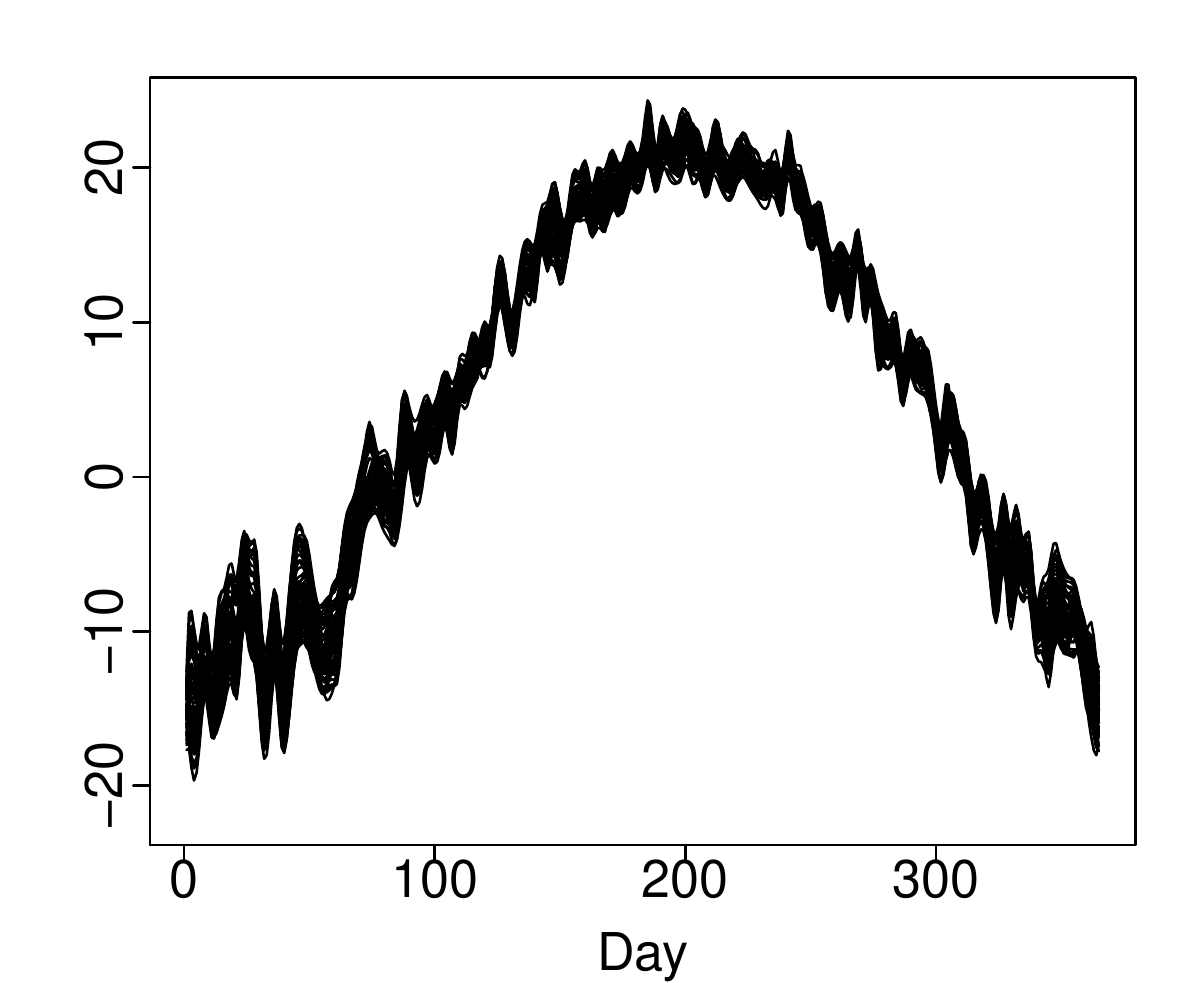}
      \includegraphics[width=5.8cm]{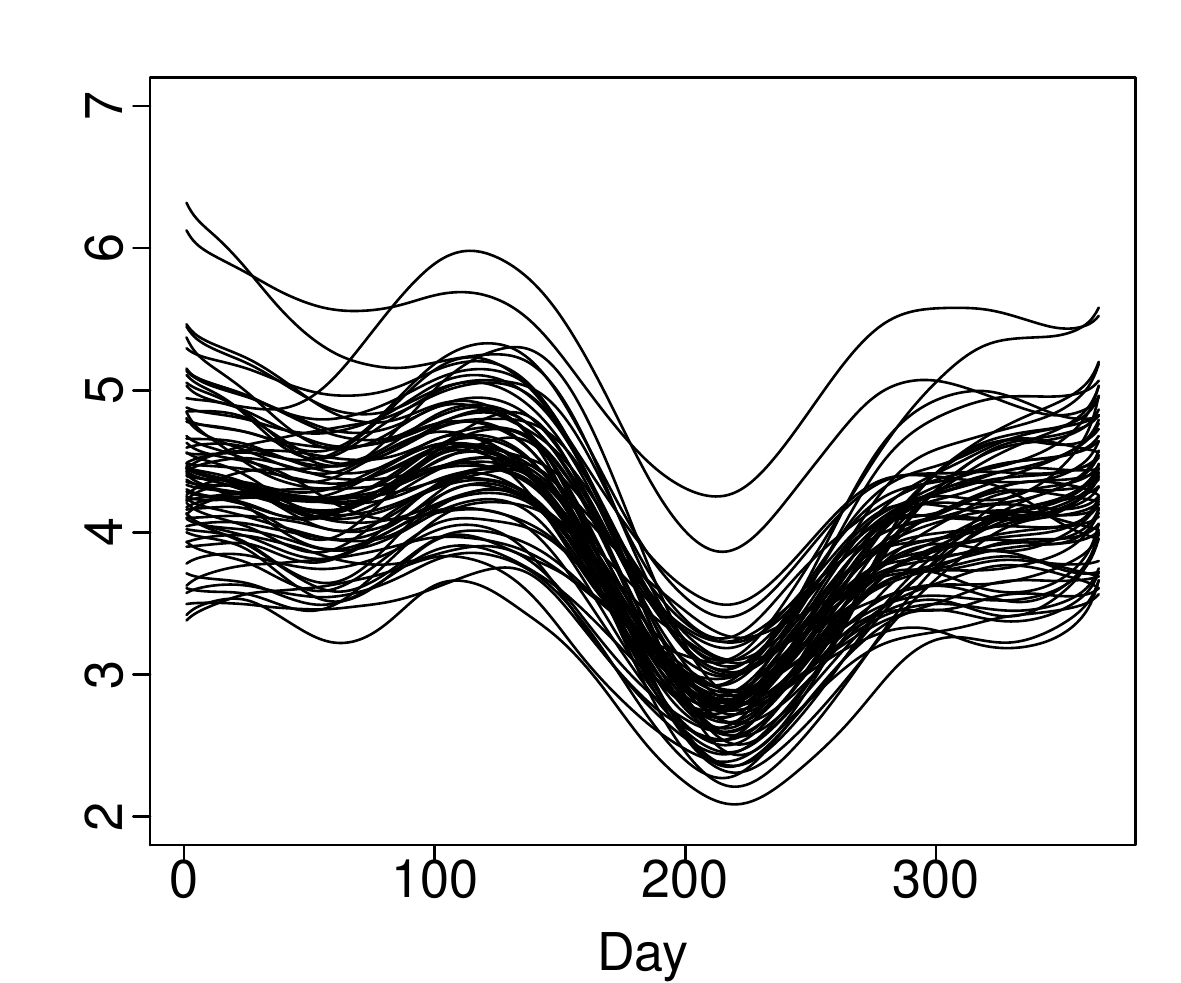}      
  \caption{Plots of discrete data (first row) and their calculated smooth functions (second row) for daily weather data.}
  \label{fig:Fig_5}
 \end{figure}

The data were averaged over the entire time, and $B$-spline basis function expansion was used to convert the discretely observed data to functional forms. Using the GCV criterion, the estimated numbers of basis functions of the temperature, wind speed, and solar radiation variables were $\left[ 147, 62, 150 \right]$. The plots of the observed dataset and its computed functions are presented in Figure~\ref{fig:Fig_5}.

\begin{figure}[!b]
  \centering
  \includegraphics[width=6cm]{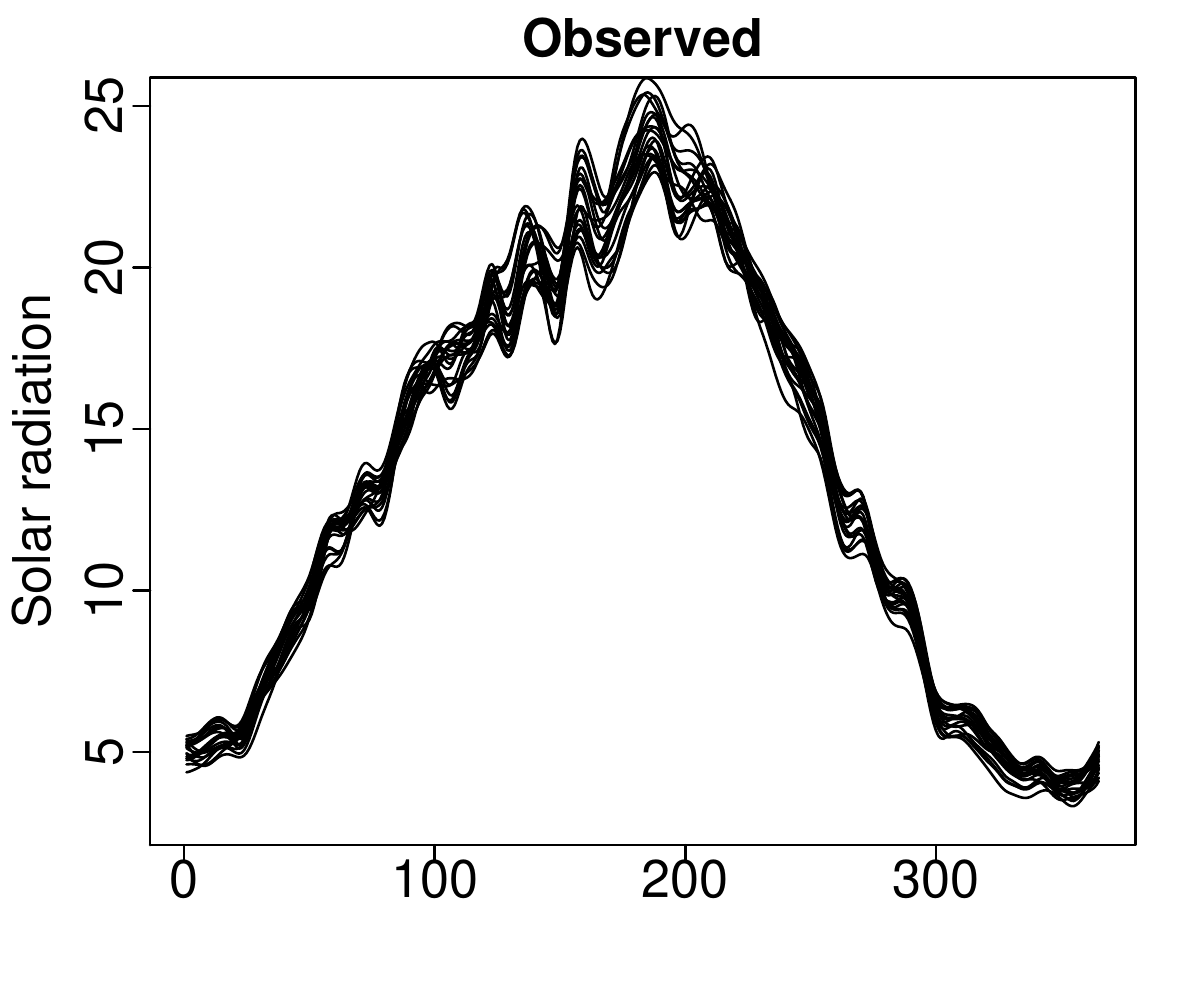}
  \quad
    \includegraphics[width=6cm]{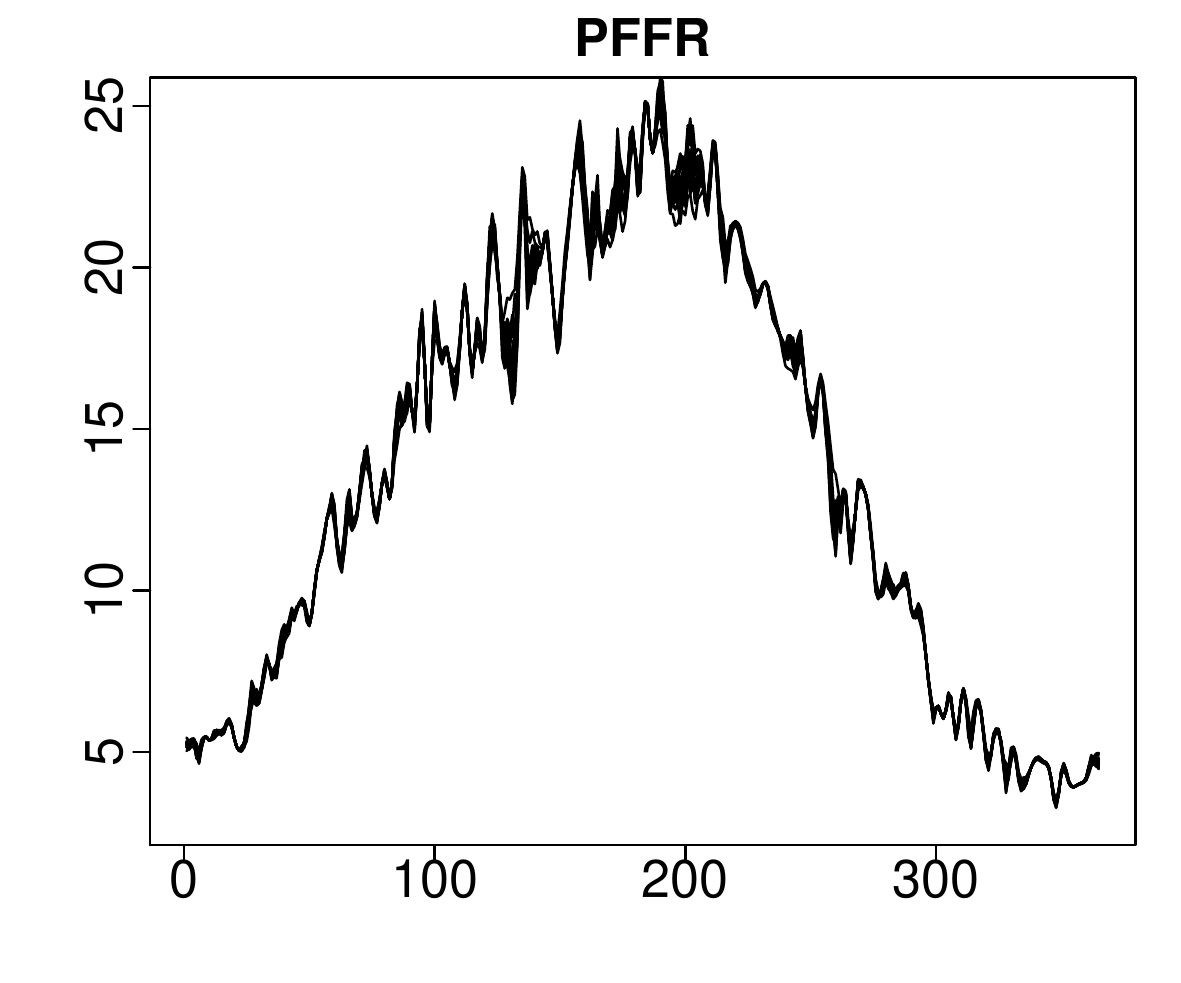}
    \\
      \includegraphics[width=6cm]{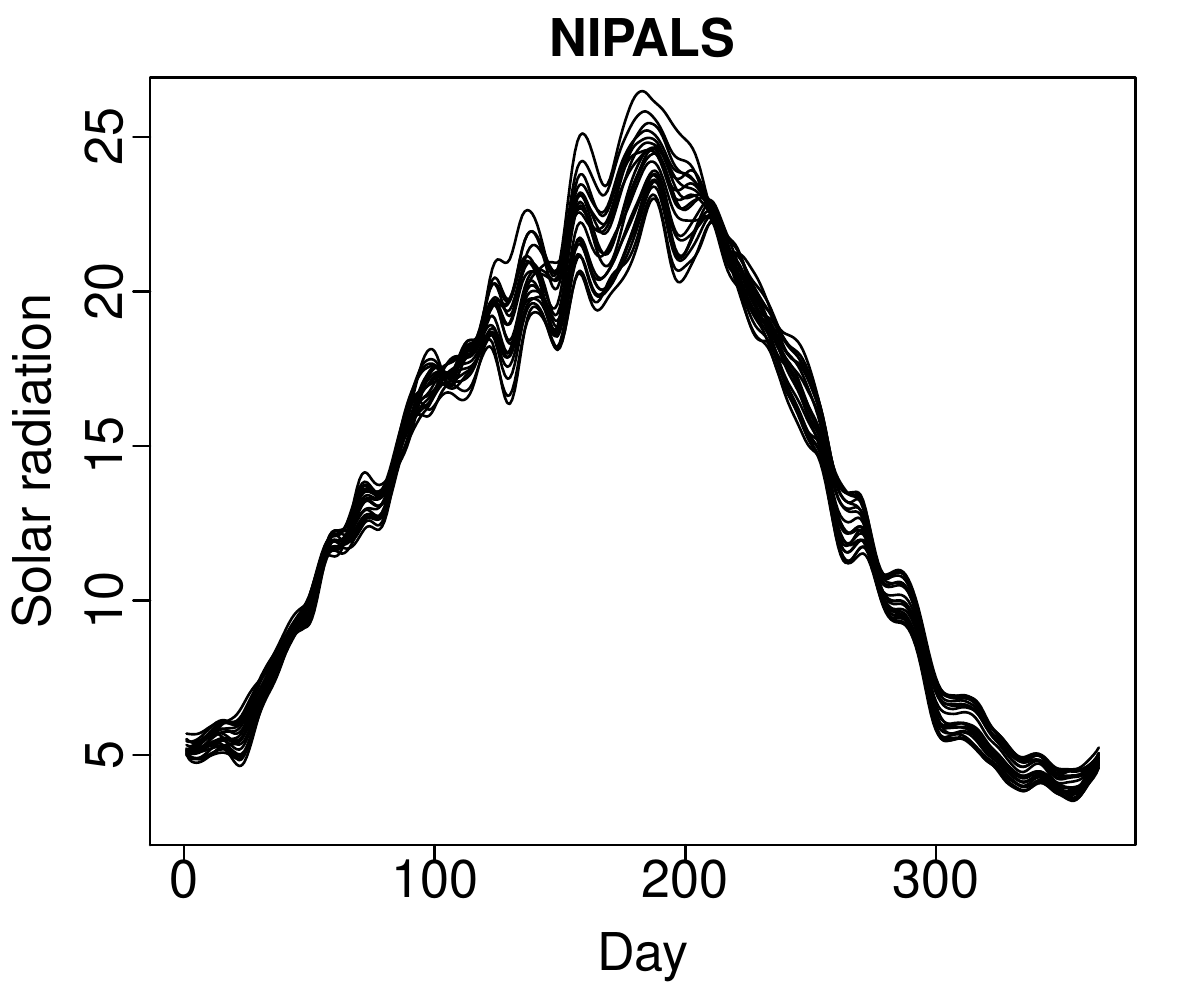}
      \quad
  \includegraphics[width=6cm]{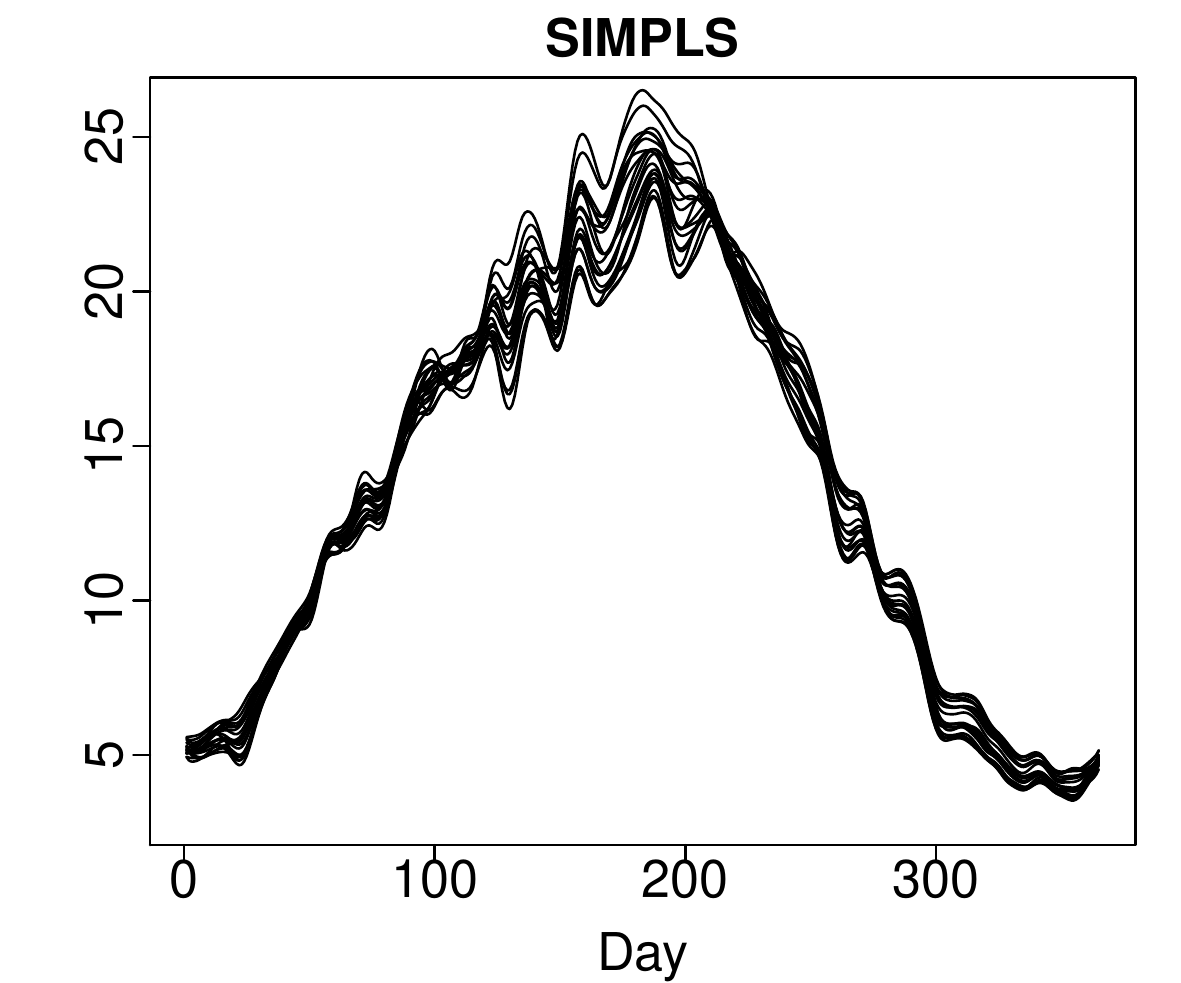}      
  \caption{Plots of observed and predicted solar radiation functions for the test stations. The FFRM was constructed using one predictor (wind speed). The PFFR, NIPALS, and SIMPLS methods were used to estimate the model parameter $\beta_1(s,t)$.}
  \label{fig:Fig_6}
\end{figure}

For the dataset, we predicted total solar radiation using temperature and wind speed variables. For this purpose, the dataset was divided into the following two parts: FFRMs were constructed based on the variables of the first 50 stations to predict the total solar radiation functions of the remaining 20 stations. However, FREG and PFFR do not allow for more than one functional predictor in the FFRM. Therefore, to compare all the methods considered in this study, we first constructed the FFRM using only one functional predictor as follows:
\begin{equation}
\Y_i(t) = \int_{S} \X_{i1}(s) \beta_1(s,t) + \epsilon_i(t) \qquad i = 1, \cdots, 50, \label{eq:simple}
\end{equation}
where $\Y_i(t)$ and $\X_{i1}(s)$ denote the $i^{th}$ function of the solar radiation and wind speed. Then, we calculated the $\text{AMSE}_p$ as follows:
\begin{equation*}
\text{AMSE}_p = \frac{1}{20} \sum_{i=51}^{70} \left[ \Y_i(t) - \widehat{\Y}^*_i(t) \right]^2,
\end{equation*}
where $\widehat{\Y}^*_i(t)$ denotes the predicted response function for $i$\textsuperscript{th} station. The MPL and FREG failed to provide an estimate for the regression parameter because of the singular matrix problem. The calculated $\text{AMSE}_p$ for the PFFR, NIPALS, and SIMPLS were $\left[ 275.0590, 100.4674, 100.2813 \right]$. The results show that, of all methods, the proposed PLS-based methods were most effective. The observed and predicted total solar radiation functions of the test stations using model~\eqref{eq:simple} are presented in Figure~\ref{fig:Fig_6}.

\begin{figure}[!htbp]
  \centering
  \includegraphics[width=5.9cm]{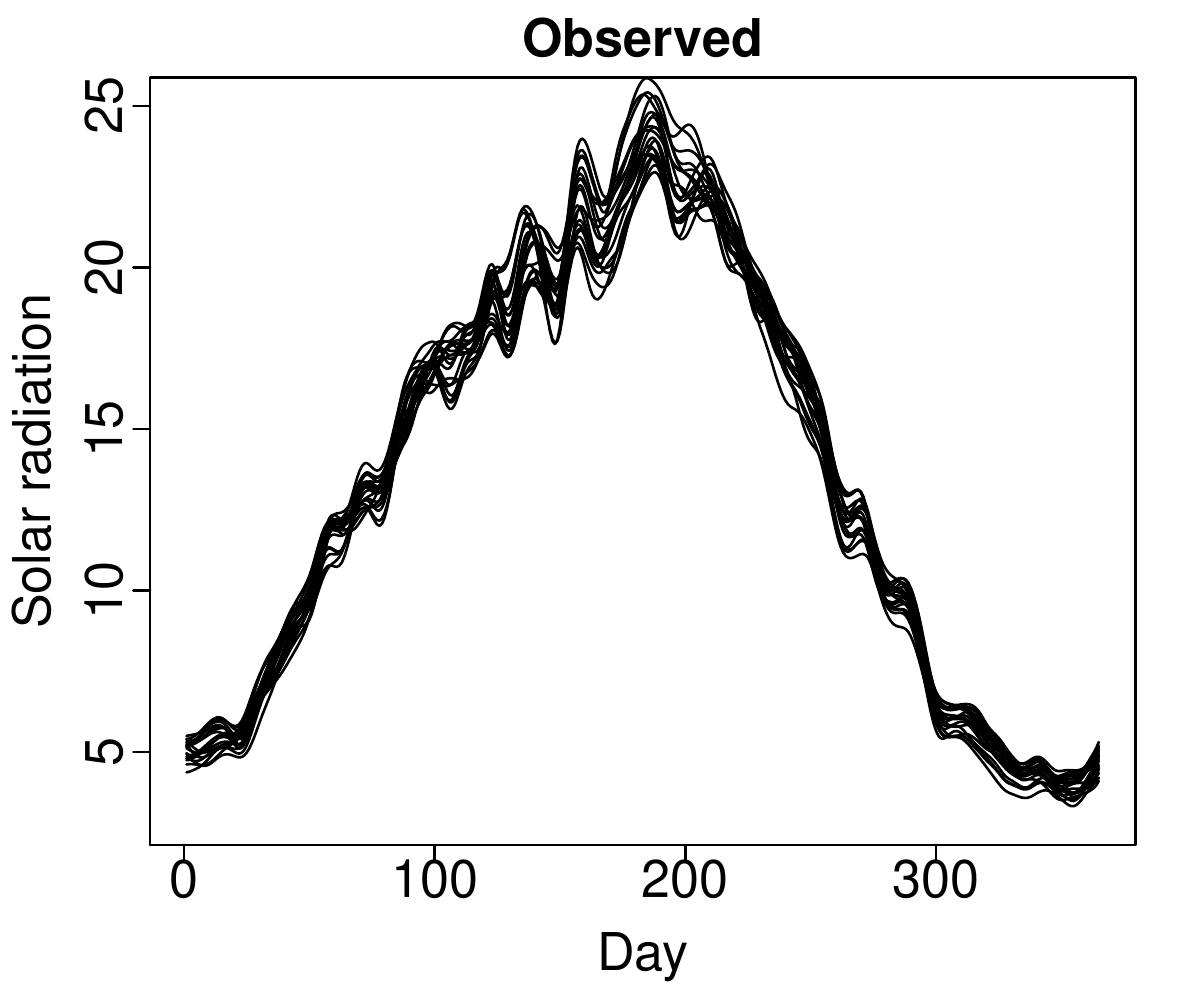}
  \includegraphics[width=5.9cm]{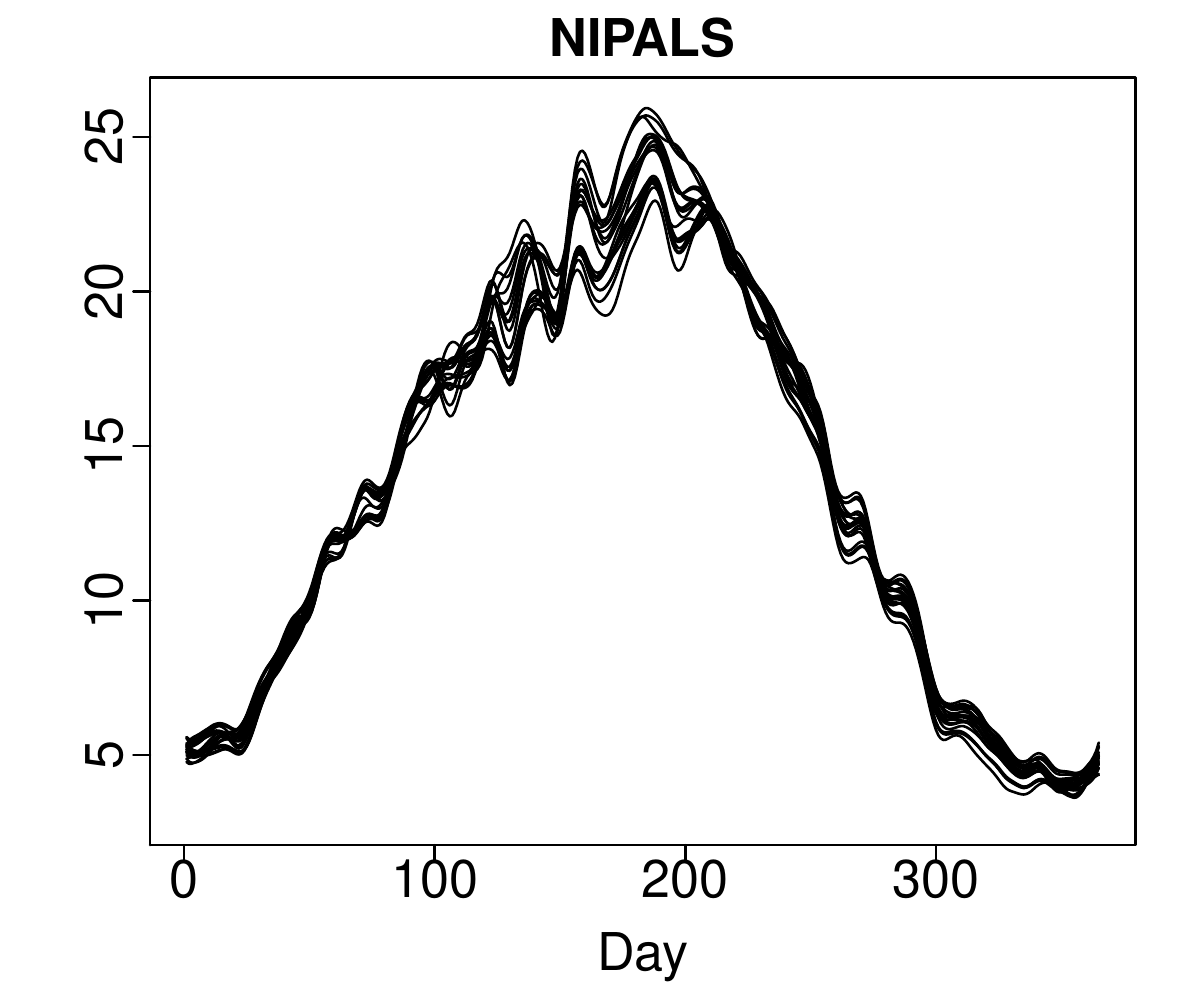}
    \includegraphics[width=5.9cm]{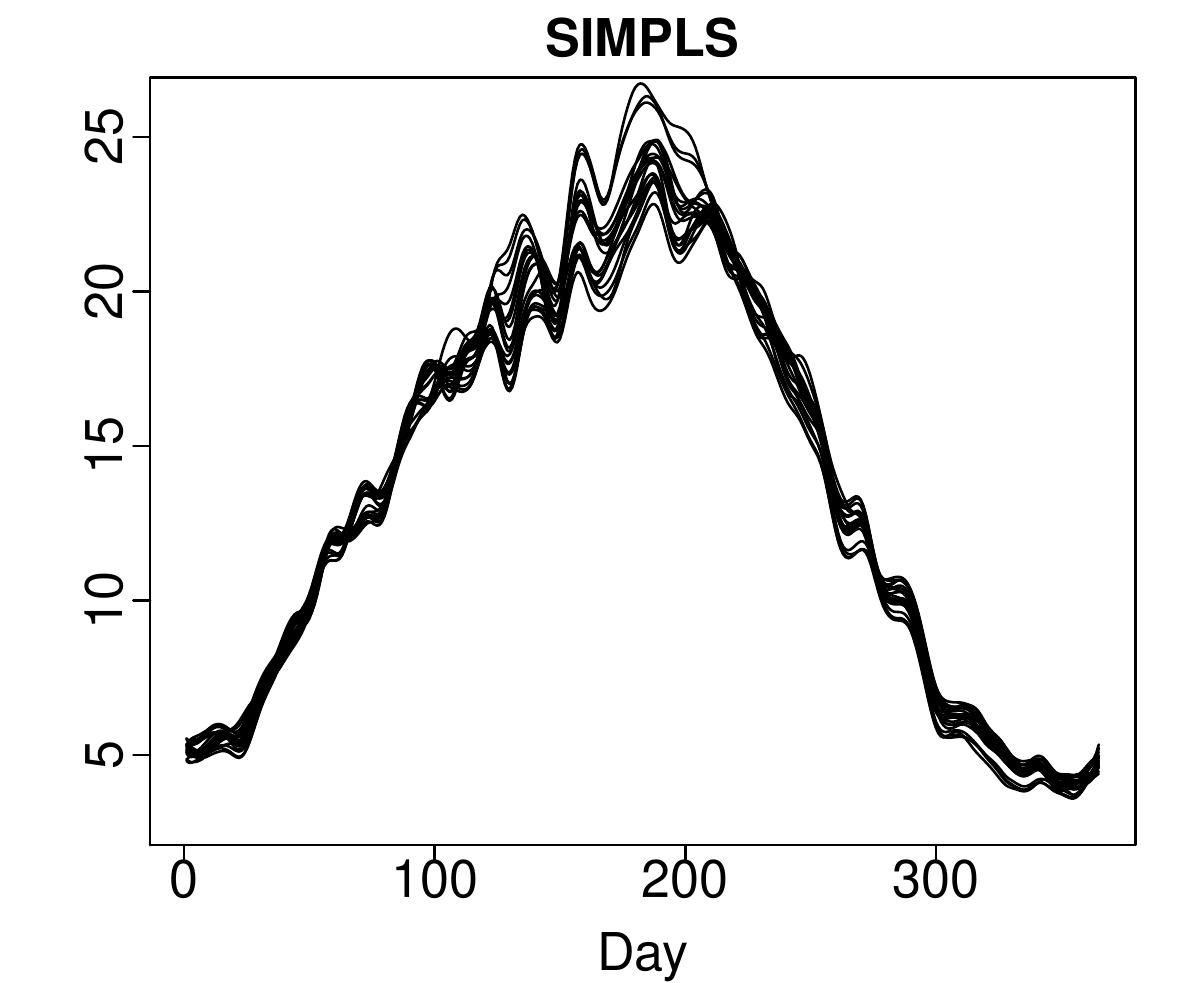}  
  \caption{Plots of observed and predicted solar radiation functions for the test stations: daily weather data. The FFRM was constructed using two predictors (wind speed and temperature). The NIPALS, and SIMPLS methods were used to estimate the model parameters $\beta_1(s,t)$ and $\beta_2(s,t)$.}
  \label{fig:Fig_7}
\end{figure}

Next, we constructed the FFRM using more than one functional predictor as follows:
\begin{equation}
\Y_i(t) = \int_S \X_{i1}(s) \beta_1(s,t) + \int_S \X_{i2}(s) \beta_2(s,t) + \epsilon_i(t), \qquad i = 1, \cdots, 50, \label{eq:multi}
\end{equation}
where $\Y_i(t)$, $\X_{i1}(s)$, and $\X_{i2}(s)$ denote the $i^{th}$ function of the solar radiation, wind speed, and temperature, respectively. In this case, we only compare the MPL, NIPALS, and SIMPLS methods because the FREG and PFFR do not allow more than one functional predictor in the model. For the data, the MPL failed to provide an estimate for the regression parameter because of the singular matrix problem; therefore, we only compared the proposed NIPALS and SIMPLS methods. The calculated $\text{AMSE}_p$ values for the NIPALS and SIMPLS, respectively, were $\left[ 56.93, 57.70 \right]$. The results show that the NIPALS performed better than the SIMPLS. The observed and predicted total solar radiation functions for model~\eqref{eq:multi} are provided in Figure~\ref{fig:Fig_7}.

In summary, our proposed PLS methods tend to produce superior performances than existing estimation methods and other available FFRMs. Additionally, the proposed methods avoided common computing problems. Computational issues observed when using the MPL and FREG are presented as follows:
\begin{verbatim}
> Error: cannot allocate vector of size 7.3 Gb
> Error in chol.default(Asym): the leading minor of order 7264 is not positive definite
> Error in solve.default(Sigma): system is computationally singular: + 
> reciprocal condition number = 1.64815e-20
\end{verbatim}
These errors were attributable to the relatively large number of basis functions estimated by the GCV. A possible solution for overcoming these problems is to use a high-performance computer or a smaller number of basis functions in the modeling phase. However, the proposed PLS-based methods can successfully provide estimates for the model parameters in a few seconds without producing any errors listed above (an example \texttt{R} code for the analysis of daily North Dakota weather data is available at \url{https://github.com/hanshang/FPLSR}).

\section{Conclusion\label{sec:conc}}

Analysis of the association between functional response and functional predictors has received considerable attention in many research fields. For this purpose, several FFRMs have been proposed, with their primary objective being to estimate the model parameters accurately. Existing estimation methods work well when a small number of predictors are used in the model. Existing estimation methods also work well when a finite number basis functions are used to convert discretely observed data to smooth functions. However, when the opposite occurs, estimation methods suffer from two key problems. First, they fail to provide estimates for the model parameters because of the singular matrix problem. Second, they are computationally time-consuming.

In the present study, we integrated the PLS approach with an FFRM and used two principal algorithms, NIPALS and SIMPLS, to estimate the parameter matrix. The finite-sample performances of the proposed approaches were evaluated using Monte-Carlo experiments and empirical data analysis. We compared our results with some other estimation methods within an FFRM. Our findings illustrate that the proposed approaches perform better than several existing estimation methods. They avoid the singular matrix problem by decomposing the response and predictor variables into orthogonal matrices. Additionally, they are computationally more efficient compared with available estimation methods.

For the proposed methods, two points need to clarify:
\begin{inparaenum}
\item[1)] Throughout this study, we assume that the functional predictor variables are observed on the same domain (see model~\eqref{Eq:FLM}). However, there may be some cases where the dataset includes multiple predictors observed on different domains \citep[see, e.g.,][]{HG18}. In such a case, the following FFRM can be considered:
\begin{equation}\label{Eq:dm}
\Y_i(t) = \beta_0(t) + \sum_{m=1}^M \int_{S_m} \X_{im}(s_m) \beta_m(s_m, t) + \epsilon_i(t),
\end{equation} 
where $S_m$ denotes the domain of $m$\textsuperscript{th} functional predictor. All the functional predictor matrices $\X_m$, for $m = 1, \cdots, M$, have the same row lengths, and they can be stacked into a vector $\pmb{\X}$. Our proposed method can also be used to estimate the variable-domain FFRM given in~\eqref{Eq:dm}. 
\item[2)] In this study, we use the same finite-dimensional basis functions method ($B$-spline) to convert the discretely observed data points of predictor variables into their functional forms. However, using different basis functions methods for different predictors may be more useful in some cases; for example, $B$-spline and Fourier bases can be used to approximate the functional variables having non-periodic and periodic structures, respectively. In such a case, the basis coefficient matrices produced by different basis expansion methods will have the same row lengths; and thus, using different basis functions for different predictors does not interfere with the use of our proposed method.
\end{inparaenum}

The present research can be extended in three directions: 
\begin{inparaenum}
\item[1)] We only considered two fundamental algorithms, NIPALS and SIMPLS, to estimate the FFRM. However, numerous algorithms, such as improved kernel PLS \citep{dayal}, Bidiag2 \citep{golub}, and non-orthogonalized scores \citep{Marthens}, are available in the PLS literature; and could be included for performance comparison.
\item[2)] In the presence of outliers, it may be advantageous to consider a robust PLS algorithm, such as the robust iteratively reweighted SIMPLS in \cite{AA17}. 
\item[3)] In our numerical analyses, the finite sample performance of the proposed method is evaluated using a fixed $h = 5$ number of PLS components. However, its performance may depend on different choices of the number of PLS components. Thus, a cross-validation approach of \cite{YT98}, \cite{Racine00}, and \cite{APS09} may be proposed to determine the optimum number of PLS components.
\end{inparaenum}

\newpage
\bibliographystyle{agsm}
\bibliography{mybibfile}

\end{document}